\documentclass[12pt]{article}
\usepackage{epsfig}
\usepackage{axodraw}

\def\bsg{\ifmmode B\to X_s\gamma\else $B\to X_s\gamma$\fi}
\def\bsll{\ifmmode B\to X_s\ell^+\ell^-\else $B\to X_s\ell^+\ell^-$\fi}
\def\shat{\ifmmode \hat{s}\else $\hat{s}$\fi}

\newcommand{\newc}{\newcommand}

\newc{\gsim}{\lower.7ex\hbox{$\;\stackrel{\textstyle>}{\sim}\;$}}
\newc{\lsim}{\lower.7ex\hbox{$\;\stackrel{\textstyle<}{\sim}\;$}}
\newc{\ie}{{\it i.e.}}
\newc{\etal}{{\it et al.}}
\newc{\mev}{\hbox{\rm\,MeV}}
\newc{\gev}{\hbox{\rm\,GeV}}
\newc{\tev}{\hbox{\rm\,TeV}}
\newc{\xpb}{\hbox{\rm\, pb}}
\newc{\xfb}{\hbox{\rm\, fb}}

\def\order#1{{\cal O}(#1)}
\newc{\mtop}{m_t}
\newc{\mbot}{m_b}
\newc{\mz}{M_Z}
\newc{\mw}{M_W}
\newc{\alphasmz}{\alpha_s(M_Z)}
\newc{\swsq}{\sin^2\theta_W}
\newc{\cwsq}{\cos^2\theta_W}
\newc{\tw}{\tan\theta_W}
\newc{\cw}{\cos\theta_W}
\newc{\sw}{\sin\theta_W}
\newc{\BR}{\hbox{\rm BR}}
\newc{\zbb}{Z\to b\bar}
\newc{\Gb}{\Gamma (Z\to b\bar b)}
\newc{\Gh}{\Gamma (Z\to \hbox{\rm hadrons})}
\newc{\sgn}{\mbox{sgn}}
\def\mpl{M_{\rm Pl}}

\def\eq#1{eq.~(\ref{#1})}
\def\fig#1{fig.~\ref{#1}}

\def\vev#1{\langle {#1} \rangle}

\def\mtil{\tilde{m}}
\def\gtilu{{\tilde g}_u}
\def\gtild{{\tilde g}_d}
\def\gtilup{{\tilde g}_u^\prime}
\def\gtildp{{\tilde g}_d^\prime}

\def\gtilupq{{\tilde g}_u^{\prime 2}}
\def\gtildpq{{\tilde g}_d^{\prime 2}}

\newlength{\myem}
\newcounter{mysubequation}[equation]

\newcommand{\TeV}{\,\mathrm{TeV}}
\newcommand{\GeV}{\,\mathrm{GeV}}
\newcommand{\MeV}{\,\mathrm{MeV}}

\newcommand{\msq}{m_{\tilde q}}
\newcommand{\msl}{m_{\tilde l}}
\newcommand{\MHc}{M_{H_C}}
\newcommand{\MAd}{M_\Sigma}

\def\beq{\begin{equation}}
\def\eeq{\end{equation}}
\def\bea{\begin{eqnarray}}
\def\eea{\end{eqnarray}}
\def\slashchar#1{\setbox0=\hbox{$#1$}           % set a box for #1
   \dimen0=\wd0                                 % and get its size
   \setbox1=\hbox{/} \dimen1=\wd1               % get size of /
   \ifdim\dimen0>\dimen1                        % #1 is bigger
      \rlap{\hbox to \dimen0{\hfil/\hfil}}      % so center / in box
      #1                                        % and print #1
   \else                                        % / is bigger
      \rlap{\hbox to \dimen1{\hfil$#1$\hfil}}   % so center #1
      /                                         % and print /
   \fi}                                         %
\catcode`@=11
\long\def\@caption#1[#2]#3{\par\addcontentsline{\csname
  ext@#1\endcsname}{#1}{\protect\numberline{\csname
  the#1\endcsname}{\ignorespaces #2}}\begingroup
    \small
    \@parboxrestore
    \@makecaption{\csname fnum@#1\endcsname}{\ignorespaces #3}\par
  \endgroup}
\catcode`@=12

\begin{document}

\baselineskip=18pt

\setcounter{footnote}{0}
\setcounter{figure}{0} \setcounter{table}{0}

\begin{titlepage}
September 2004 \hspace*{\fill} CERN-PH-TH/2004-183 \\ 
% \hspace*{\fill} hep-ph/0409232

\begin{center}
\vspace{1cm}

{\Large \bf

Aspects of Split Supersymmetry}

\vspace{0.8cm}

{\bf N. Arkani-Hamed$^1$, S. Dimopoulos$^2$, G.F. Giudice$^3$, A.
Romanino$^3$}

\vspace{.5cm}

{\it $^1$ Jefferson Laboratory of Physics, Harvard University,\\
Cambridge, Massachusetts 02138, USA}

{\it $^2$ Physics Department, Stanford University, \\ Stanford,
California 94305, USA}

{\it $^3$ CERN, Theory Division, \\ CH-1211 Geneva 23,
Switzerland}

\end{center}
\vspace{1cm}

\begin{abstract}
\medskip
We explore some fundamental differences in the phenomenology,
cosmology and model building of Split Supersymmetry compared with traditional
low-scale supersymmetry. We show how the mass spectrum of Split
Supersymmetry naturally emerges from theories where the dominant
source of supersymmetry breaking preserves an $R$ symmetry,
%$D$-breaking of global supersymmetry leads to
%a split spectrum, where squarks and sleptons are ultraheavy, whereas
%gauginos and higgsinos are at the TeV scale
characterize the class of theories where the unavoidable $R$-breaking
by gravity can be neglected, and point out a new possibility, where
supersymmetry breaking is directly communicated at tree level to the
visible sector via renormalizable interactions. Next, we discuss
possible low-energy signals for Split Supersymmetry. The absence of
new light scalars removes all the phenomenological difficulties of
low-energy supersymmetry, associated with one-loop flavor and CP
violating effects. However, the electric dipole moments of leptons and
quarks do arise at two loops, and are automatically at the level of
present limits with no need for small phases, making them accessible
to several ongoing new-generation experiments. We also study proton
decay in the context of Split Supersymmetry, and point out scenarios
where the dimension-six induced decays may be observable. Finally, we
show that the novel spectrum of Split Supersymmetry opens up new
possibilities for the generation of dark matter, as the decays of
ultraheavy gravitinos in the early universe typically increase the
abundance of the lightest neutralino above its usual freeze-out value.
This allows for lighter gauginos and Higgsinos, more accessible both
to the LHC and to dark-matter detection experiments.
%decay rates from model-independent dimension 6 operators can be
%within the range of proposed upgrades, but only if the
%gauginos-higgsinos weigh $\sim 10 TeV$, or there are favorable
%threshold effects.

\end{abstract}

\bigskip
\bigskip

%% \begin{flushleft}
%% June 2004
%% \end{flushleft}

\end{titlepage}

%%%%%%%%%%%%%%%%%%%%%%%%%%%%%%%%%%%%%%%%%%%%%%%%%%%%%%%%%%%%%%%
%\tableofcontents
%\vfill\eject

\section{Introduction}

Our concept of Naturalness -- the principle that Nature abhors
fine-tunings -- is based on theories with a few vacua. It has led
to the proposal of low-scale supersymmetry, in order to avoid
tuning short-distance parameters to 30-decimals. But this
principle faces a serious challenge from the cosmological constant
problem, where we see no new physics at the scale $\sim 10^{-3}$
eV required by naturalness.

Recent developments in string theory~\cite{string} suggest the
existence of an enormous ``landscape'' of long-lived metastable
vacua. These can have a significant impact on the criterion of
naturalness, and may provide the key to resolving the cosmological
constant problem along the lines of ref.~\cite{Weinberg}. They may also
have a bearing on the solution to the hierarchy problem. For
example, if the number of vacua that  break supersymmetry at
high-scale is more than $10^{30}$ times larger than those with
low-scale supersymmetry, then the breaking of supersymmetry at
high scales is  favored~\cite{Giryavets:2004zr}. The 30-decimal
tuning is compensated by the enormous ``entropy factor'' favoring
high-scale breaking of supersymmetry. In this case, the simplest
possibility would be that the Standard Model (SM) is preferred to
the usual low-energy supersymmetric SM~\cite{georgi}. This would
then account for why we have not seen any evidence for
supersymmetry -- either in the spectrum or in rare process, such
as FCNCs, CP violation, proton decay, etc. --  at the expense of
giving up two major successes of the supersymmetric SM: gauge
coupling unification~\cite{stuart} and natural dark-matter
candidate~\cite{gold}. A more interesting  possibility that
preserves these successes  is that  approximate chiral symmetries
protect the fermions of the supersymmetric SM  down to the TeV
scale~\cite{savnim,noi}. So, the sparticle spectrum of these
theories is ``split" in two: {\it (i)} the scalars (squarks and
sleptons) that get a mass at the high-scale of supersymmetry
breaking $\mtil$, which can be as large as the GUT scale, and {\it
(ii)} the fermions (gauginos and higgsinos) which remain near the
electroweak scale and can account for both gauge-coupling
unification and DM. The only light scalar in this theory is a
finely-tuned Higgs. So, rather than the dull prediction that the
LHC will discover just the Higgs, these theories -- coined Split
Supersymmetry -- predict gauginos and higgsinos at a TeV, maintain
the successes of the supersymmetric SM, and account for the
absence of problematic flavor and  CP-violation, of fast proton
decay, and of an excessively light Higgs, caused by the presence
of light squarks and sleptons in the supersymmetric SM.

In this paper we address some novel theoretical, cosmological and
phenomenological aspects of Split Supersymmetry\footnote{Some
consequences of Split Supersymmetry have been considered in
ref.~\cite{tutti}.}. An important theoretical issue is how the
spectrum of Split Supersymmetry can naturally emerge from a
high-energy theory. In sects.~\ref{patt} and \ref{anomed} we
characterize the general class of theories that have this
property. We also present in sect.~\ref{direct} models where the
supersymmetry breaking is directly mediated at tree level to the
visible sector of the theory.

%to construct theories in which the Split-Supersymmetry
%spectrum  emerges naturally,
%and in  section 2 and 3 we characterize such a general class. We show that
%theories in which supersymmetry is broken by
%what we characterize as ``$D$-breaking"
%naturally have the split spectrum.

Having abandoned naturalness, the crucial ingredient pinning the
gaugino and Higgsino masses to the TeV scale is the requirement for
the lightest supersymmetric particle (LSP) to form the dark matter of
the universe. Novel properties of the gravitino in Split
Supersymmetry can impact this analysis. In sect.~\ref{grasec} we
discuss the properties of gravitinos in Split Supersymmetry,
emphasizing their impact on the issue of dark matter. Gravitino decay
generates a non-thermal population of dark-matter neutralinos, which
gives an additional contribution to the energy density on top of the
usual freeze-out abundance. This, in general, suggests that the mass
of the supersymmetric particles can be lighter than what derived from
the thermal abundances -- making Split Supersymmetry potentially more
accessible to the LHC and to dark-matter detection experiments.

%  Section 4 deals with the cosmology and DM abundance in these theories. A
%novel possibility is that the now heavy scalar sparticles can decay or
%annihilate in the early universe into gravitinos. Since the gravitinos do not
%annihilate efficiently, this results in an overabundance of gravitinos which
%subsequently, after the
%gravitinos decay, result in an overabundance of the LSP
%DM, compared to the usual freezout abundance. This, in general, suggests that
%the mass of the LSP is lighter than that of  the naive freezout calculation -
%making the light sparticles potentially more accessible to the LHC.

While the decisive verdict on Split Supersymmetry awaits TeV scale
collider experiments, it is important to examine what signals can show
up in other, low-energy experiments. Having eliminated the new
scalars, the rate for anomalous effects in FCNC processes, which are
troublesome for usual low-energy supersymmetry, are predicted to be
unobservably small.  Similarly, the proton decay rates from dimension
5 operators become irrelevant -- these problems are eliminated so
efficiently that there are no associated signals to be seen. However,
in sect.~\ref{edmsec} we show that in Split Supersymmetry, the
contribution to lepton and quark electric dipole moments (EDM),
arising at two loops, are naturally of the order of the current experimental
limits for CP violating phases of order unity. This is to be
contrasted with usual low-energy supersymmetry, where CP violation is
a problematic feature, with electric dipole moments arising at
one-loop being too large by roughly a factor of $10^2$--$10^3$ for
soft-term phases of order one. The naturally observable EDM are an
exciting generic prediction of Split Supersymmetry, and may be checked
by a new generation of experiments that plan to improve the limit by a
factor of at least $\sim 100 - 1000$, possibly before the turn-on of
the LHC. We also study proton decay in Split Supersymmetry in
sect.~\ref{prosec}, and point out scenarios under which the
dimension-6 mediated decays may be observable.

In sect.~\ref{concsec} we conclude with a detailed summary of our
results and remarks on directions for future research.

% We subsequently discuss proton decay. The contribution of dimension five
%operators to p-decay is supressed, since the squarks are heavy. Dimension six
%operatorsare potentially intereting because Split
%Supersymmetry is closer to the
%non-supersymmetric standard model than
%low-scale supersymmetry, and the unification
%scale is somewhat smaller. This is in part compensated by the larger value of
%$\alpha_{GUT}$ and the resulting proton decay rate is still too small for the
%planned upgrades of p-decay experiments - unless threshold effects are
%favorable.
%  The last phenomenological issue of interest is the case when the
%gravitino is the LSP. This gives rise to the interesting signature of the
%long-lived decay of the gluino into gravitino and a gluon jet.

%****Somewhere in the introduction I should mention direct mediation, I am not
%sure where it will go. It will roughly be as follows:******
%A new theoretical possibility of split supersymmetry
%is that supersymmetry breaking is
%directly felt, through renormalizable couplings, to the visible sector, and
%resuts in heavy scalar sparticles...

\section{Split Supersymmetry and the Pattern of Supersymmetry Breaking}
\label{patt}

The spectrum of Split Supersymmetry is defined by squark masses
and $B_\mu$ term generated at a large mass scale $\mtil$, and by
gaugino masses and $\mu$ term of the order of the weak scale. In
this section we want to characterize the conditions under which
the mechanism of supersymmetry breaking leads to such a spectrum,
rather than the usual case in which all supersymmetric partners
are approximately mass-degenerate. The question is reminiscent of
the $\mu$ problem: why isn't $\mu = \order{\mpl}$ generated in the
exactly supersymmetric theory? In Split Supersymmetry we are
encountering an extended $\mu$ problem: why aren't gaugino masses
and $\mu = \order{\mtil}$ generated in the broken supersymmetric
theory?

At first sight, it may appear that the ordinary case of
mass-degenerate superpartners is the most generic since, after
supersymmetry is broken, all these particles are expected to
acquire masses. We want to show that this is not necessarily the
case. The nature of the mass spectrum depends on the presence or
the absence of an approximate (or accidental) $R$-symmetry of the
observable sector of the theory. Once this symmetry is realized,
the mass spectrum of Split Supersymmetry naturally emerges.

It is known~\cite{nelson} that there is a connection between the
existence of $R$-symmetries in the hidden sector and of
supersymmetry breaking. If the relevant superpotential is a {\it
generic} function of fields (\ie\ all interaction terms are
present with no tuned coefficients), the presence of an
$R$-symmetry is a necessary condition for supersymmetry breaking,
while a spontaneously broken $R$-symmetry provides a sufficient
condition. However, because of the non-renormalization theorem in
supersymmetry, it is not unusual to encounter superpotentials that
are {\it non-generic} functions of fields, where certain
interaction terms are absent. Many example of such
superpotentials, leading to spontaneously-broken supersymmetry,
are known. Moreover, $R$-symmetry cannot be an exact symmetry once
the theory is extended to supergravity and the cancellation of the
cosmological constant is allowed. Indeed, such cancellation is
obtained by tuning the gravitino mass to a value proportional to
the scalar component of the superpotential $W$. Since $R[W]=2$,
the condition of vanishing cosmological constant necessarily leads
to a breaking of the $R$-symmetry.

Let us turn to consider the various patterns of supersymmetry
breaking, paying special attention to the properties of the
underlying $R$-symmetry.

\subsection{F-breaking}

We start by reviewing standard results regarding the origin of the
soft terms, in order to introduce notations and to allow a
comparison with the analysis presented in the next section.

Supersymmetry breaking is parametrized by a spurion chiral
superfield \beq X=1 +\theta^2 \mtil . \eeq The superfield $X$
breaks supersymmetry and $R$-symmetry, since both its scalar and
auxiliary components have a background value. Masses for the
scalar components of visible-sector superfields $Q$ ($\mtil_Q$),
gaugino masses ($M_{\tilde g}$), and trilinear $A$-terms are
generated by the operators \bea \int d^4\theta X^\dagger X
Q^\dagger Q &\to & \mtil_Q^2=\mtil^2,
\label{op1}\\
\int d^2\theta X W_\alpha W_\alpha &\to & M_{\tilde g}=\mtil ,
\label{op2}\\
\int d^2\theta X Q^3 &\to & A=\mtil . \label{op3} \eea Here
$W_\alpha$ is the gauge superfield strength, and $Q^3$ is a
gauge-invariant combination of visible-sector superfields. All
soft terms are of the order of the mass scale $\mtil$.  Notice
that, in the absence of supersymmetry breaking ($\mtil \to 0$),
the interactions in eqs.~(\ref{op1})--(\ref{op3}) are
$R$-invariant, once we assign $R[X]=0$. Therefore, supersymmetry
and $R$-symmetry are simultaneously broken by the auxiliary
component $F_X=\mtil$.

For phenomenological reasons, we have to exclude the operator \beq
\int d^2\theta M_* H_1H_2, \label{eccomu} \eeq where $H_{1,2}$ are
the two Higgs-doublet superfields and $M_*$ is a mass scale
characteristic of the exact-supersymmetry theory. This can be
achieved by: {\it (i)} a global $U(1)$ $PQ$ symmetry with
$PQ[H_1H_2]\ne 0$; or {\it (ii)} an $R$-symmetry with
$R[H_1H_2]\ne 2$; or {\it (iii)} a mechanism that does not allow
to embed this term in the underlying GUT. Once this is done,
supersymmetry breaking generates the $\mu$-term and $B_\mu$ (the
scalar counterpart of the $\mu$-term) through the
operators~\cite{giud} \bea \int d^4\theta X^\dagger  H_1H_2 &\to &
\mu =\mtil ,
\label{op4}\\
\int d^4\theta X^\dagger X H_1H_2 &\to & B_\mu =\mtil^2,~\mu
=\mtil . \label{op5} \eea These interactions are allowed, in the
supersymmetric limit, by a combination of the $R$ and $PQ$
symmetries such that the $R$ charges are $R[H_1H_2]=0$.
Consistently, this symmetry forbids the unwanted operator in
\eq{eccomu}. In conclusion, all possible soft terms (including
$\mu$) have been generated at the scale $\mtil$.

\subsection{D-breaking}

Next, we want to examine the case in which supersymmetry breaking does
not lead to $R$-symmetry breaking. Therefore we consider the spurion
superfield \beq Y=1+\theta^4 \mtil^2, \label{dspur} \eeq which breaks
supersymmetry, but preserves an $R$-symmetry if we assign $R[Y]=0$.
This spurion could arise from a genuine supersymmetry $D$-breaking in
the gauge sector. However, it could also be induced by a chiral
superfield $X$, with $\vev{X}=0$ and $\vev{F_X}\ne 0$. As an example,
this happens in the simplest of all theories of supersymmetry
breaking, with a linear superpotential for a chiral superfield $X$
\begin{equation}
W = \mu^2 X \,.
\end{equation}
This has $F_X = \mu^2$ and broken supersymmetry. The vev of $X$ is
undetermined at this level; however, if there are higher dimension
operators in the K\"ahler potential of the form
\begin{equation}
\delta K = -\frac{(X^\dagger X)^2}{M^2} \,,
\end{equation}
then $X$ is stabilized at the origin $\langle X \rangle = 0$, with
$m_X^2 = \mu^4/M^2 > 0$. (In fact, precisely this sort of term in the
K\"ahler potential is generated at 1-loop in renormalizable
O'Raifeartaigh models of supersymmetry breaking). If such a field $X$
appears in the interactions with visible-sector fields only in the
combination $X^\dagger X$, then the visible sector will have
supersymmetry breaking without $R$-breaking. In particular, this
happens if $X$ is a non-singlet under a hidden-sector gauge group (as
is often the case in models with dynamical supersymmetry breaking) or
if there exists a discrete symmetry $X\to -X$. For simplicity, we will
call ``$D$-breaking'' the case characterized by \eq{dspur}.  However,
it should be clear that we do not necessarily require that
supersymmetry breaking is triggered by a gauge superfield, but rather
that the supersymmetry breaking is not accompanied by $R$ breaking.

Couplings of the spurion $Y$ to the visible sector induce the soft
terms \bea \int d^4\theta Y Q^\dagger Q &\to & \mtil_Q^2=\mtil^2,
\label{op6}\\
\int d^4\theta Y H_1H_2 &\to & B_\mu =\mtil^2 . \label{op7} \eea
No other renormalizable operator is allowed. Notice that we have
defined our spurions as dimensionless, and so we are effectively
working to all orders in $Y$. Only visible-sector fields determine
the operator dimensionality.

Since we have excluded the operator in \eq{eccomu}, possibly by a
GUT mechanism, the visible sector has an accidental $R$-symmetry
with $R[Y]=0$ and $R[H_1H_2]=0$, even after supersymmetry is
broken. This is the symmetry that forbids the appearance of
gaugino masses, $A$ and $\mu$ terms. This symmetry is accidental,
since it has not been imposed on the theory, but it is just the
consequence of supersymmetry, gauge symmetry, field
dimensionality, and the absence of the operator in \eq{eccomu},
once $D$-breaking is assumed. Notice, in particular, that imposing
on the theory an exact $R$-symmetry is not justified since, as we
have previously discussed, the cancellation of the cosmological
constant necessarily implies an explicit $R$ breaking.

The situation is analogous to how lepton number is implemented in
the SM. Lepton number is only an accidental SM symmetry and it is
expected to be violated in the fundamental (possibly GUT) theory by
effects of order unity. The smallness of the neutrino mass is then
explained not by small coefficients, but by the large scale
hierarchy between the electroweak and GUT masses. Similarly, in
$D$-breaking, even if the $R$-symmetry is broken in the
fundamental theory by effects of order unity, gaugino masses and
the $\mu$ parameter will be suppressed with respect to squark
masses and $B_\mu$, whenever there is a certain hierarchy between
the fundamental (possibly Planck) scale and $\mtil$.

Equations~(\ref{op6}) and (\ref{op7}) allow for the possibility of a
cancellation of the Higgs mass term. In particular, a K\"ahler
structure of the kind \beq \int d^4\theta Y \left( \sin\beta H_1
  -\cos\beta H_2^\dagger \right) \left( \sin\beta H_1^\dagger
  -\cos\beta H_2 \right) , \label{kabo} \eeq leads to a massless Higgs
superfield, for any value of $\beta$. Although the cancellation needs
a fine-tuning, it is interesting that Split Supersymmetry allows this
possibility for the Higgs mass and not, for instance, for squark
masses, so there can be a reason why of all the scalars of the
supersymmetric SM, only the Higgs is fine-tuned to be light.

%If the structure in \eq{kabo} were exact after radiative
%corrections, and violated only by the dynamics that generate $M_{\tilde
%g}$ and $\mu$, one could relate gaugino and higgsino masses to the
%weak scale, independently of the dark-matter argument. In reality,
%an unjustified tuning is necessary. Nevertheless, it is interesting to
%remark that Split Supersymmetry allows for the possibility of just two
%(unexplained) tunings: cosmological constant and Higgs mass. No analogous
%cancellation is possible for squark and slepton masses.

To generate the other soft terms, we have to include the leading
non-renormalizable operators, \bea \frac{1}{M_*}\int d^4\theta Y
W_\alpha W_\alpha r&\to & M_{\tilde g}=
\frac{\mtil^2}{M_*}r ,\label{op8}\\
\frac{1}{M_*}\int d^4\theta Y Q^3 r &\to & A=
\frac{\mtil^2}{M_*}r ,\label{op9}\\
\frac{1}{M_*}\int d^4\theta Y D^2 (H_1H_2) r &\to & \mu =
\frac{\mtil^2}{M_*}r .\label{op10} \eea Here $M_*$ is the mass
that characterizes the interactions between hidden and visible
sectors. In supergravity, we identify $M_*$ with $\mpl$, but
smaller values of $M_*$ are possible in alternative schemes of
supersymmetry-breaking mediation. The mass spectrum of Split
Supersymmetry naturally emerges from $D$-breaking, whenever $\mtil
\ll M_*$. The parameter $r$ that we have included in
eqs.~(\ref{op8})--(\ref{op10}) measures a possible further
suppression caused by an (approximate) $R$-symmetry present in the
dynamics that mediates supersymmetry breaking at the scale $M_*$.
It could be viewed as a scalar spurion field carrying an $R$
charge. Indeed, if $R[r]=-2$, the interactions in
eqs.~(\ref{op8})--(\ref{op10}) are $R$ invariant. If the
underlying dynamics does not preserve an $R$-symmetry, we can set
$r=\order{1}$.

Next, we can establish what is the reasonable minimum value of $r$
in theories where there is an energy range for which the 4-D
supergravity description is valid. This is given by the effect of
gravitational interactions that communicate to the visible sector
the unavoidable $R$ violation coming from the cancellation of the
cosmological constant. In supergravity, the positive vacuum energy
associated with supersymmetry breaking is cancelled by having a non-zero
expectation value for the superpotential $W$ which contributes $-3
|W|^2$ to the vacuum energy. As $W$ has $R$-charge 2, this breaks
$R$, and gives a gravitino mass 
\beq 
m_{3/2} = e^{\frac{K}{2
\mpl^2}}\frac{|W|}{\mpl^2} \,. 
\eeq 
Here $\mpl =2.4\times 10^{18}$~GeV
is the reduced Planck mass. We can get a reasonable lower bound on
$m_{3/2}$ as follows: given scalar masses of order $\tilde{m}$,
there are contributions to the vacuum energy of at least order
$\sim \tilde{m}^4/(16 \pi^2)$, and therefore we expect at
least
\begin{equation}
\frac{|W|^2}{\mpl^2} \gsim \frac{\tilde{m}^4}{16 \pi^2}
\quad\Rightarrow\quad m_{3/2} \gsim \frac{\tilde{m}^2}{4 \pi \mpl} \,.
\end{equation}
Now, in order to generate gaugino masses, we must break both $R$ and
supersymmetry. The minimum amount of $R$-breaking will come from the
non-zero $W$. Local operators of the form
\begin{equation}
\int d^4 \theta \frac{W_\alpha W_\alpha W^\dagger
\left(Q^\dagger Q + \cdots \right)}{M_*^6} \,,
\end{equation}
where $Q$ are generic chiral fields in the visible sector, can
not be protected by any symmetries, and indeed we expect
gravitational loops to generate such operators as counterterms
with $M_* \sim \mpl$. Now, only given the supersymmetry breaking of the
heavy scalar masses $\tilde{m}$, we can expect that the terms
$Q^\dagger Q + \cdots$ have a non-vanishing $\theta^2
\bar{\theta}^2$ component of size comparable to the vacuum energy
of the low-energy theory beneath $\tilde{m}$, which is $\sim
\frac{\tilde{m}^4}{16 \pi^2}$ , leading to
\begin{equation}
\label{gaucc}
M_{\tilde{g}} \gsim \frac{\tilde{m}^6}{\mpl^5} \sim
\frac{m_{3/2}^3}{\mpl^2} \,,
\end{equation}
ignoring loop factors. Barring cancellations, this can be taken as
a plausible minimum value of the gaugino mass,  since it is the
result of just the existence of gravity and the cancellation of
the cosmological constant. This magnitude can be estimated
diagrammatically in terms of a one-loop gravitational contribution
to gaugino masses (induced by graviton/gaugino and
gravitino/gauge-boson exchange), which is cutoff in the
ultraviolet by the supersymmetry-breaking mass $m_{3/2}$. The
phenomenological requirement that \eq{gaucc} does not exceed the
TeV gives the constraint $m_{3/2} \lsim 10^{13}$ GeV. 
Incidentally, we remark that in the case of gravity mediation ($\mtil
\sim m_{3/2}$) this bound is numerically equivalent to the one
derived by the condition that the gluino lifetime is shorter than
the age of the universe, which is necessary to evade the experimental
limits from searches of anomalously heavy isotopes~\cite{savnim}.
Finally, notice that
considerations similar to those we have made for gaugino masses
also apply to the $\mu$ term. A Higgs-gravitino
loop with a $B_\mu$ mass insertion induces $\mu \sim
m_{3/2}^3/\mpl^2$, taking the gravity-mediation result $B_\mu \sim
m_{3/2}^2$.

We have seen that there is some connection between the
gaugino/Higgsino and gravitino masses, as they both break $R$. Is
there any relation between $\tilde{m}$ and $m_{3/2}$? In fact
these can usefully be thought of as the independent parameters
characterizing a model of Split Supersymmetry. There are theories where
$\tilde{m} \ll m_{3/2}$ -- this occurred in one of the models in
ref.~\cite{savnim}, where the gravitino mass $m_{3/2}$ came from
Scherk-Schwartz breaking with a fifth dimension as $m_{3/2} \sim
1/R$, while the scalar masses were $\tilde{m} \sim \frac{1}{R^2
\mpl}$. Using the fifth dimension to lower the higher-dimensional
Planck scale to the GUT scale actually predicted $m_{3/2} \sim 1/R
\sim 10^{13}$ GeV, gaugino/Higgsinos right at the TeV
scale, and $\tilde{m} \sim 10^{9}$ GeV. There are also theories
in which $\tilde{m} \gg m_{3/2}$, as in the models of direct
mediation we discuss in sect.~\ref{direct}.

\section{Contributions from Anomaly Mediation}
\label{anomed}

We have argued that the breaking of $R$ required by the
cancellation of the vacuum energy in supergravity will eventually
infect the visible sector, generating gaugino masses of order at
least $M_{\tilde g} \sim m_{3/2}^3/\mpl^2$. However, there is a
potentially larger effect, coming from anomaly
mediation~\cite{anmly}, which typically generates
$M_{\tilde g} \sim \alpha/(4 \pi) m_{3/2}$. For $m_{3/2} \lsim
100$ TeV or so, this will be subdominant to whatever generates the
gaugino masses, but what happens for $m_{3/2} \gg M_{\tilde g} $?
In ref.~\cite{savnim}, a concrete model of supersymmetry breaking using a
fifth dimension was constructed, where the anomaly-mediated
contributions were small, with $m_{3/2} \sim 10^{13}$ GeV, in fact
predicting gaugino and Higgsino masses near $\sim 100$ GeV. Our
purpose here is to explore the issue in more general terms: why do
we get the anomaly-mediated contribution, and under what
circumstances is it naturally suppressed?

Let us review the origin of the anomaly-mediated contribution to
gaugino masses
\begin{equation}
M_{\tilde g} = \frac{\beta(g)}{g} F_\phi \,,
\end{equation}
where $\beta(g) =dg/d\ln\mu$ and $\phi = 1 + \theta^2 F_{\phi}$ is the
conformal compensator, which appears in the Lagrangian in form
\begin{equation}
{\cal L} = \int d^4 \theta \phi^\dagger \phi K +\left( \int d^2
  \theta \phi^3 W +\mbox{h.c.} \right) \,.
\end{equation}

In order to see what the induced $F_\phi$ is, we set $K,W$ equal to
their expectation values in the supersymmetry-breaking background (with $K|_0 =
-3\mpl^2$ to canonically normalize the Einstein action), and in
components we find that the potential is
\begin{equation}
V_{\rm vac} = 3 \mpl^2 |F_{\phi}|^2 + (F_{\phi}^* A + \mbox{h.c.}) + V \,.
\end{equation}
Here
\begin{equation}
A = -3 W^*|_{0} - K|_{\theta^2}, \quad V = -K|_{\theta^4} -
2\mbox{Re}\, W|_{\theta^2}\,,
\end{equation}
and the gravitino mass is
\begin{equation}
|m_{3/2}| = \frac{|W|}{\mpl^2} \,.
\end{equation}
The equation of motion for $F_\phi$ now simply determines
\begin{equation}
\label{fphi}
F_\phi = -\frac{A}{3 \mpl^2} = m_{3/2} + \frac{K|_{\theta^2}}{3
\mpl^2} \,,
\end{equation}
and this gives the vacuum energy
\begin{equation}
V_{\rm vac} = V - \frac{|A|^2}{3 \mpl^2} \,.
\end{equation}
Fine-tuning the vacuum energy to zero, we also find that
\begin{equation}
\label{dir}
|F_\phi| = \sqrt{\frac{V}{3 \mpl^2}} \,.
\end{equation}
In theories in which supersymmetry is broken dynamically without
singlets in the hidden sector, the term $K|_{\theta^2}/(3\mpl^2)$ in
eq.~(\ref{fphi}) is negligible, and we obtain $|F_\phi| =
m_{3/2}$. In general, however, $F_\phi$ is not proportional to
$m_{3/2}$, but it is directly determined by $V$ as in eq.~(\ref{dir}). 
% Note that $F_{\phi}$ is directly determined by $V$ and more
% indirectly by $m_{3/2}$.

In a theory where supersymmetry is broken already in the global limit, $V$
can be thought of as the ``vacuum energy'' of the supersymmetry breaking
sector. More precisely, it is the Goldstino decay constant; the
Lagrangian for the Goldstino $\chi$ has a quadratic part
\begin{equation}
V \bar{\chi} \bar{\sigma}^\mu \partial_\mu \chi + \cdots \,.
\end{equation}
Thus in theories where supersymmetry breaking is decoupled from gravity, so
that $V$ and $K|_{\theta^2}$ remain constant as $\mpl \to \infty$, we
have $|F_\phi| = m_{3/2}$. However, this is not the case for theories
where supersymmetry breaking is intimately tied to gravity and supersymmetry is restored
as $\mpl \to \infty$, so that $V \to 0$ in this limit. This may appear
at first sight to be a sick limit, since the coefficient of the
Goldstino kinetic term is going to zero.  But in fact, in such models,
the Goldstino does get a kinetic term via mixing with the gravitino.
This is the secret of the no-scale supergravity structure, and also
helps explain why the no-scale structure is so ubiquitous in
compactifications of theories with extra dimensions. As is familiar,
the scalar radius moduli $T$ only acquire kinetic terms by mixing with
gravity, and the same is true for their fermionic partners $\psi_T$.
Indeed, we can see this easily in a theory with a single extra
dimension \cite{lutyokada} with a radius modulus $T = r + \theta
\psi_T + \cdots$. Since $\mpl^2 = 2M_5^3 {\rm Re}(r)$ we have
\begin{equation}
K = -3 M_5^3(T + T^\dagger)
\end{equation}
and we can add a constant superpotential
\begin{equation}
W = M_5^3 c \,.
\end{equation}
It is easy to see that $V=0$ and $F_\phi$ = 0, since both
$K|_{\theta^4}$ and $W|_{\theta^2}$ vanish, while
\begin{equation}
F_T = c^*, \quad m_{3/2} = \frac{c M_5^3}{\mpl^2}
\end{equation}
and supersymmetry is broken. This is equivalent \cite{kaplanweiner} to a
Scherk-Schwarz breaking with $c$ setting the phase picked up by the
gravitino traversing the dimension.

At this level the modulus $r$ is undetermined; in any case the
K\"ahler potential will be modified by quantum corrections, and we may
also have additional terms in the superpotential. Suppose then that
the K\"ahler and superpotentials are modified by introducing the
functions $k(T + T^\dagger)$ and $w(T)$ as follows
\begin{equation}
K = -3 M_5^3 (T + T^\dagger) + k(T + T^\dagger), \, W = c M_5^3 +
w(T) .
\end{equation}
Let us work to first order in $k,w$. 
By computing the equation of motion for $F_T$, we find
\begin{equation}
F_\phi = \frac{c^* k^{\prime \prime} + w^{\prime *}}{3 M_5^3} =
\frac{|c|^2 k^{\prime \prime} + w^{\prime *} c}{3m_{3/2} \mpl^2} .
\end{equation}
While the potential is
\begin{equation}
V_{\rm vac}(T) = V_{K} + V_{W} ,
\end{equation}
where
\begin{equation}
V_K = -|c|^2 k^{\prime \prime}, \, V_W =   -2 \mbox{Re}(w^{\prime}
c^*) .
\end{equation}
Note that
\begin{equation}
- \mbox{Re}\left( \frac{F_\phi}{m^*_{3/2}}\right) 
= \frac{V_K + \frac{1}{2} V_W}{3
|m_{3/2}|^2 \mpl^2} = \frac{V_{\rm vac}(T) + V_K}{6 |m_{3/2}|^2 \mpl^2} \to
\frac{V_K}{6 |m_{3/2}|^2 \mpl^2} ,
\end{equation}
where in the last expression we have put in the fine-tuning to make
$V_{\rm vac}(T)$ vanish at the minimum of the potential. From here we can
conclude that
\begin{equation}
\left|\frac{F_\phi}{m_{3/2}}\right| \geq
\left| \mbox{Re}\left( \frac{F_\phi}{m^*_{3/2}}\right) \right| 
= \frac{V_K}{6 |m_{3/2}|^2
\mpl^2}
\label{disgz}
\end{equation}

The inequality above is saturated when there are no relative phases
between $F_\phi$ and $m_{3/2}^*$. We typically expect that the real
and imaginary parts of $F_\phi$ are comparable, so the right hand
side of the above is a good estimate for the actual size of
$|F_\phi |$.

What can we expect for the range of possible natural values for
$V_K$ and thus $|F_\phi|$? Clearly, $V_K$ gets a contribution at
least from the one-loop vacuum energy of the non-supersymmetric theory beneath
the gravitino mass. Barring cancellations, this yields a
reasonable lower bound of
\begin{equation}
V_{K min} \sim \frac{m_{3/2}^4}{16 \pi^2} \,.
\end{equation}
An upper bound on $V_K$ is of order $\sim m_{3/2}^2 \mpl^2$, and
therefore we can bound
\begin{equation}
\frac{m_{3/2}^3}{16 \pi^2 \mpl^2} \lsim |F_\phi| \lsim m_{3/2} \,.
\end{equation}
The lower bound will be saturated in any theory where supersymmetry breaking
shuts off above the mass of the gravitino. An example is provided by
the model of ref.~\cite{savnim}, where the gravitino mass is about $1/r$
and the scale above which supersymmetry breaking shuts off is also $1/r$,
with the radius stabilized by balancing Casimir energies scaling as
$1/r^4 \sim m_{3/2}^4$. Note that, when this lower bound is saturated, the
anomaly-mediated contribution to gaugino masses is subdominant to
the direct gravitational contribution we discussed in the previous
section.

The expression for $F_\phi$ in \eq{disgz} has a nice
physical interpretation. The Goldstino field here is the fermionic
component of $T$; defining $\psi_T = \langle F_T \rangle \chi = c^*
\chi$ (to first order), under a supersymmetry transformation $\delta_\zeta 
\chi =
\zeta$. The quadratic part of the Goldstino kinetic term is then, to
leading order,
\begin{equation}
-|c|^2 k^{\prime \prime} \bar{\chi} \bar{\sigma}^\mu \partial_\mu
\chi = V_K \bar{\chi}\bar{\sigma}^\mu \partial_\mu \chi
\end{equation}
so we can identify $V_K$ with the Goldstino decay constant.  The link
between $F_\phi$ and the Goldstino decay constant is not an accident,
and it is illuminating to understand its origin without any reference
to the conformal compensator formalism, directly at component level.
Suppose we have somehow broken supersymmetry and cancelled the vacuum energy to
get flat space. The gravitino $\psi_\mu$ is massive and its quadratic
action is of the form
\begin{equation}
\mpl^2 \left(\epsilon^{\mu \nu \alpha \beta} \bar{\psi}_{\mu}
\bar{\sigma}_\nu
\partial_\alpha \psi_\beta + m_{3/2} \psi_\mu \sigma^{\mu \nu}
\psi_\nu \right) \,.
\end{equation}
Of course, the massive gravitino has extra degrees of freedom -- the
longitudinal polarizations -- compared to the massless gravitino. This
is reflected in the fact that the above Lagrangian is no longer
invariant under the gauge transformation $\delta_\zeta \psi_\mu =
\partial_\mu \zeta$ that were used to reduce the degrees of freedom
described by $\psi_\mu$ to the two transverse polarizations. As
familiar in ordinary gauge theories, it is useful to explicitly
introduce the longitudinal modes. This can simply be done by
performing the broken-gauge transformation and promoting the
transformation parameter to a field. For massive gauge bosons, for
instance, we have the familiar
\begin{equation}
f^2 A_\mu A^\mu \to f^2 (A_\mu + \partial_\mu \theta)^2 \supset
f^2 (\partial_\mu \theta)^2
\end{equation}
and we see that the gauge boson mass term turns into the kinetic term
of the ``eaten" longitudinal mode $\theta$, which shifts under the
gauge symmetry.

Doing the same thing for the gravitino, we have
\begin{equation}
m_{3/2} \psi_\mu \sigma^{\mu \nu} \psi_\nu \to m_{3/2}
(\partial_\mu \chi + \psi_\mu) \sigma^{\mu \nu} (\partial_\nu \chi
+ \psi_\nu) \,.
\end{equation}
Already we notice a significant difference with the gauge theory case
-- the only possible kinetic term for the Goldstino $\chi$ appears to
be of the form $(\partial \chi)^2$, which is sick for a fermion.
However, due to the antisymmetry of $\sigma^{\mu \nu}$, this kinetic
term actually vanishes (which is in fact what dictates the use of
$\sigma^{\mu \nu}$ rather than $\eta^{\mu \nu}$ in the gravitino mass
term). The Goldstino here does indeed get a healthy kinetic term, {\it
  but only by mixing with gravity}.  We can now couple the gravitino
to all the other fields of the theory, including all massive regulator
fields, in the standard way, and at this level there is no induced
supersymmetry breaking anywhere else in the spectrum. This analysis is however
puzzling if we think of a theory that breaks supergravity already in
the globally supersymmetric limit. There, the Goldstino exists
independently of gravity! How do we describe this situation? After
all, in the unitary gauge we still have a massive gravitino described
by the same effective Lagrangian.

The resolution is that the gravitino transformation property under
supersymmetry is changed to
\begin{equation}
\delta \psi_\mu =\partial_\mu \zeta + i f \sigma_\mu \bar{\zeta}
\,.
\end{equation}
Before saying anything about the size of $f$, we can identify $f$
with $F_\phi$ in the conformal compensator formalism. Because of the
modified transformation law for $\psi_\mu$, if we now couple the
gravitino to other fields, what used to be supersymmetric will no
longer be so, and the variation can only be compensated by adding
supersymmetry breaking terms to the matter Lagrangian from the outset. It is
easy to see that the new term in the variation of $\psi_\mu$ is
simply the superpartner of a conformal transformation (compare
$\delta h_{\mu \nu} = \eta_{\mu \nu} \phi$), and the required change
of the Lagrangian is conveniently made by turning on $F_\phi = f$ in
a conformal compensator field.

With the modified transformation, in promoting $\zeta$ to the
Goldstino field $\chi$, we find the quadratic kinetic term for
$\chi$
\begin{equation}
V_K \bar{\chi} \bar{\sigma}^\mu
\partial_\mu \chi + \cdots \,,
\end{equation}
where the Goldstino decay constant $V_K$ is
\begin{equation}
V_K = |f|^2 \mpl^2 + 2 \, \mbox{Re}(m_{3/2} \mpl^2 f) =
\mpl^2\left(|f + m_{3/2}^*|^2 - |m_{3/2}|^2\right) \,.
\end{equation}

Inverting this, we can bound $|f/m_{3/2}|$ in terms of $V_K$ as
\begin{equation}
\left|\frac{f}{m_{3/2}} \right| \geq \sqrt{1 + \frac{V_K}{m_{3/2}^2
\mpl^2}} - 1 \,.
\label{above}
\end{equation}

The above expression then bounds $F_{\phi}=f$ directly in
terms of physical quantities: the Goldstino decay constant $V_K$ and
the gravitino mass $m_{3/2}$. Again in theories when all the
relevant parameters are real, the inequality above becomes an
equality.

Note as before that in theories in which supersymmetry is already
broken in the limit of global supersymmetry, $V_K = 3 m_{3/2}^2
\mpl^2$ and \eq{above} becomes
\begin{equation}
\frac{|F_\phi|}{m_{3/2}} \to 1 \,.
\end{equation}
On the other hand, in theories where the Goldstino dominantly gets
its kinetic term by mixing with the gravitino, $V_K \ll m_{3/2}^2
\mpl^2$, and \eq{above} gives
\begin{equation}
\frac{|F_\phi|}{m_{3/2}} \to \frac{V_{g}}{2 m_{3/2}^2 \mpl^2} \ll
1 \,.
\end{equation}

Summarizing, then, we have a simple characterization of the size of
anomaly mediation: in any theory where supersymmetry is broken by
non-gravitational dynamics, so that the supersymmetry breaking survives
taking $\mpl \to \infty$, anomaly mediation will persist, but in
theories where supersymmetry breaking is tied to gravity and the Goldstino
gets it kinetic term by mixing with gravity, $F_\phi$ and anomaly
mediation are suppressed.

Even when $F_\phi$ is comparable to $m_{3/2}$, the supersymmetric SM
fields can be insulated from anomaly-mediated soft masses. An
example \cite{Luty} is provided by the theories where the SM fields
descend from a broken CFT or, in an AdS/CFT dual language, live on an
IR brane of a warped compactification. There is now a new dynamical
field -- the dilaton S of the broken CFT -- and the K\"ahler and
superpotentials for the SM superfields have the structure
\begin{equation}
K = e^{S + S^\dagger} \hat{K}, \quad W = e^{3 S} \hat{W} \,.
\end{equation}
Including the conformal compensator, the Lagrangian has the form
\begin{equation}
\int d^4 \theta \left( -3 \mpl^2 \phi^\dagger \phi + \phi^\dagger
\phi \, e^{S + S^\dagger} \hat{K}\right) +\left( \int d^2 \theta \phi^3
e^{3S} \hat{W} +\mbox{h.c.} \right)\,.
\end{equation}
We can now define a new field $\omega = \phi e^S$, which then
decouples $\phi$ from direct communication from the SM sector
\begin{equation}
\int d^4 \theta \left( -3 \mpl^2 \phi^\dagger \phi + \omega^\dagger
\omega \hat{K}\right) + \left( \int d^2\theta \omega^3 \hat{W}
+\mbox{h.c.} \right) \,. 
\end{equation}
Now, even breaking supersymmetry in some generic way generating $F_\phi$, there
is no anomaly-mediated contribution to the soft terms. Of course again
the dilaton $\omega$ must be stabilized, but the form of the
stabilization and the resulting size of the supersymmetry breaking proportional
to $F_\phi$ are model-dependent. It is clear that we can use this
mechanism to arbitrarily shield the visible sector from the breaking of
{\it supergravity}, which is why ref.~\cite{Luty} dubbed this
observation ``supersymmetry without supergravity". In particular, we
can again consistently making $m_{3/2} \gg M_{\tilde g}$.

\section{Direct Mediation of Supersymmetry Breaking}
\label{direct}

One of the ubiquitous features of usual theories of supersymmetry
breaking is the presence of a hidden sector where supersymmetry is
broken, and only indirectly mediated to the visible fields. It is
obviously simpler to imagine that the supersymmetric SM fields pick up a mass
directly at tree-level, by renormalizable interactions with a
supersymmetry breaking sector.  However, this hope was dashed very
early on. First, with only $F$-term breaking, the only supersymmetry
breaking masses for chiral superfields are $B_\mu$ type terms, which
makes it impossible to make all the scalars heavy -- in particular
there is always a scalar lighter than the up
quark~\cite{Ferrara:1979wa}. The addition of new gauge factors opens
up the possibility of using $D$-terms. However, there is a bigger
problem: tree-level breaking can generate scalar masses, but not
gaugino masses, which will then necessarily be suppressed at least by
a loop factor relative to the scalars. This forces the scalars so
heavy that a tuning of at least $\sim 10^{-3}$ is re-introduced for
electroweak symmetry breaking.

The situation is clearly different in Split Supersymmetry, where we
are allowed to consider gauginos much lighter than the heavy scalars.
Thus, the possibility of direct mediation can be re-examined in this
context. The goal is to find a theory where squarks and sleptons
pick up a mass directly at tree level from the supersymmetry breaking
sector. If the supersymmetry breaking preserves an $R$-symmetry, the
gauginos and Higgsinos can only acquire masses from higher dimension
operators suppressed by a scale $M_*$. As long as $M_*$ is
sufficiently smaller than $\mpl$, all the physics associated with
supergravity and in particular the gravitational breaking of the
$R$-symmetry are irrelevant; in such models, the gravitino mass can
naturally be comparable or even smaller than gaugino and Higgsino
masses.

We can illustrate this general idea with a simple model. Since we want
to generate $m^2 Q^\dagger Q$ type masses for the supersymmetric SM scalars at
tree-level, the SM fields must be charged under a new gauge symmetry
that gets a $D$-term. So, let us begin with the simplest example of a
supersymmetry breaking theory of this form. Consider a $U(1)$ gauge
symmetry and chiral superfields $X,Z,\phi,\phi^c,Y,Y^c$ with $U(1)$
charges $0,0,-1,1,-1,1$ respectively, together with other possible
fields $\psi_i$ with charges $q_i$. The superpotential contains
\begin{equation}
W = \lambda X (\phi \phi^c - m^{\prime^2}) + m \phi^c Y +
\lambda^\prime \phi Y^c Z
\end{equation}
and $m,m^\prime$ are some comparable mass scales. This theory
spontaneously breaks supersymmetry: the $Y$ equation of motion forces $\phi^c = 0$,
which is in conflict with the $X$ equation and D-flatness for the
$U(1)$. It is easy to analyze the physics in a limit where we
imagine $m^\prime \gg m$. The $U(1)$ gauge symmetry is broken and
we have a vector multiplet of mass $g m^\prime$, and $Y^c,Z$ also
pair up to get a mass $\lambda^\prime m^\prime$, while the
superpotential for $Y$ is
\begin{equation}
W = m^\prime m Y \,,
\end{equation}
clearly breaking supersymmetry. Integrating out the massive vector
multiplet at tree-level also generates new terms in the K\"ahler
potential for $Y$ and the $\psi_i$.  It is easiest to see this in
superspace: ignoring the $1/g^2 W_\alpha^2$ kinetic term for the
gauge field, and going to unitary gauge, we have in the K\"ahler
potential
\begin{equation}
K = m^{\prime 2} (V^2 + \cdots) + Y^\dagger (1 + V + \cdots) Y +
\psi_i^\dagger (1 + q_i V + \cdots) \psi_i \,,
\end{equation}
and we can trivially integrate out $V$ to obtain the correction
\begin{equation}
-\frac{Y^\dagger Y}{2 m^{\prime 2}} \left(\frac{Y^\dagger Y}{2} +
q_i \psi_i^\dagger \psi_i\right) \,.
\end{equation}
The superpotential forces
\begin{equation}
F_Y = m m^\prime
\end{equation}
and this in turn generates soft masses for the $Y$ and $\psi_i$
from the corrected K\"ahler potential
\begin{equation}
m^2_Y = \frac{m^2}{2} , ~~ \quad m^2_{\psi_i} = \frac{q_i m^2}{2} \,.
\end{equation}
Of course these are nothing but the $D$-term contributions to the
soft masses of these charged fields, although it is more
appropriate to simply integrate out the heavy field, rather than
using a description with
the vev of a heavy multiplet. Note
importantly that $m_Y^2 > 0$, so that $Y$ is stabilized at $Y =
0$, preserving an $R$-symmetry with the obvious charges. Also,
note that if we had any couplings of $X$ to other fields, say of
the form
\begin{equation}
W = h X H_u H_d
\end{equation}
then there are induced couplings to $Y$ in the low-energy
effective theory, which can be seen simply by noting that this
coupling can be shifted into a redefinition of $m$ as $m^2 \to m^2
+ h/\lambda H_u H_d$, and so we have
\begin{equation}
W = m m^\prime Y \to \sqrt{m^{\prime 2} + H_u H_d} m Y = m
m^\prime Y + \frac{m}{2m^\prime} Y H_u H_d \,.
\end{equation}
The fields $H_u,H_d$ then pick up a $B_\mu$ soft term of the order
of
\begin{equation}
B_\mu = \frac{m^2}{2} \sim m_{\psi_i}^2 \,.
\end{equation}
This can be simply thought of as arising from a non-vanishing
$F_X$.

We now wish to directly couple this sector to the supersymmetric 
SM. Clearly, we
cannot only have the SM matter: because of anomalies these would
have to have both positive and negative charge, and some of the
soft masses would be negative. Also, the usual SM Yukawa
couplings would need to be made invariant. There are a number of
ways to proceed at this point, but a simple possibility is have SM
fields $f$ and Higgses $H_{u,d}$ uncharged under the $U(1)$, but
mixing with a vector like copy of all these fields $(F,F^c),
(\tilde{H}_{u,d,},\tilde{H}_{u,d}^c)$, where the
$(\tilde{H}_{u,d})$ fields are completed into $(5 + \bar{5})$'s.
These fields have masses of order $\mu$ and Yukawa couplings of
the form $f \phi^c F^c$, so that after the $U(1)$ breaking, the
light fields are an $O(1)$ mixture of the neutral fields and those
of charge $+1$, all of which have positive soft masses of order
$\sim \mu^2$. Also, the $B_\mu$ term for the Higgses arises from
the coupling $X H_u H_d$ which, as we have seen, is naturally of
the same order of magnitude. Thus, we have found
\begin{equation}
\tilde{m}^2 \sim B_\mu \sim m^2 \,. \label{confus}
\end{equation}
At this level the gauginos and Higgsinos are massless. These in
turn can be induced by integrating out physics at a higher scale
$M_*$, perhaps near the GUT scale, which can give rise to
couplings of the form
\begin{equation}
\int d^2 \theta \frac{X}{M_*} W^\alpha W^\alpha, \int d^4 \theta
\frac{X}{M_*} H_u H_d \,,
\end{equation}
which generates
\begin{equation}
M_{\tilde g} \sim \mu \sim \frac{\tilde{m}^2}{M_*} \,.
\end{equation}
Finally, the fermionic component of $Y$ is the Goldstino, which
will be eaten by the gravitino to pick up a mass of order
\begin{equation}
m_{3/2} \sim \frac{\tilde{m}^2}{\mpl}
\end{equation}
Meanwhile, without any particular suppression, the anomaly
mediated contribution to the gaugino masses is of order $M_{\tilde
g}^{\rm
  anom} \sim \alpha/(4 \pi) (\tilde{m}^2/\mpl)$, so that even for $M_*
\sim \mpl$ the contributions we have calculated dominate the
gravitational ones. In these theories, the gravitino can naturally
be comparable in mass or lighter than the gauginos and Higgsinos
(the phenomenological consequences will be studied in
sect.~\ref{lightgravitino}). If we wish $M_*$ to be above the GUT
scale, we have $\tilde{m} \gsim 10^9$ GeV.

Note also that about this same scale $\mtil$ we have all the extra
vector-like matter, affecting the running of the SM couplings
above $\tilde{m}$. In our simple example, we have added an
equivalent of 14 $(5 + \bar{5})$'s at $\tilde{m}$. Even with the
addition of all these new states, the SM couplings do not blow up
beneath the GUT scale; however, the value of the GUT coupling can
become so large that the rate from dimension-6 induced proton
decay can become significant. The existence of extra vector-like
matter charged under the SM is a feature of all direct mediation
models, though the precise size of the sector is model-dependent.
Therefore, we do generically expect the $p$-decay rates to be
significantly enhanced in these models. Further discussion of
proton decay in Split Supersymmetry will be given in section 7.

\section{Dark Matter and Gravitinos in Split Supersymmetry}
\label{grasec}

In Split Supersymmetry the all-important mass scale of
gauginos-higgsinos is decoupled from the electroweak scale and is set
solely by the requirement of obtaining the correct dark matter
abundance. It is therefore crucial to carefully compute the DM
abundance in these theories.

A fundamentally new feature of  Split Supersymmetry is that the masses of
squarks and sleptons, as well as the gravitino, can be much
heavier than a TeV. This can lead to new processes contributing to
the DM abundance. In particular, the heavy scalars can decay or
annihilate into gravitinos which hover, till they eventually decay
into a final state including the LSP. Since this process occurs in
addition to the canonical Lee-Weinberg annihilation, it typically
results in an LSP number density exceeding the usual freezout
abundance, and therefore in a lighter and more accessible LSP.

Another possibility is that the gravitino is lighter than all the
gauginos-higgsinos, and therefore it is the LSP. This leads to
interesting laboratory signatures in the decays of the NLSP and the
gluino, summarized in sect.~\ref{lightgravitino}.
% It turns out that
% the gaugino-higgsino masses, even in this case, have typically
% $\sim TeV$ upper limit, and can never exceed $\sim ~100 TeV$.

We will discriminate the various possible scenaria according to
the gravitino mass which, as we have discussed in
sect.~\ref{patt}, is an important parameter  characterizing the
theory. Let us start with some general properties of gravitinos.
The gravitino decay width is given by \beq \Gamma_{3/2}=\left(
N_g+\frac{N_m}{12}\right) \frac{m_{3/2}^3}{32\pi \mpl^2} =\left(
\frac{m_{3/2}}{10^5\gev }\right)^3 32~{\rm sec}^{-1}. \eeq Here
$N_g$ and $N_m$ are the numbers of available decay channels into
gauge--gaugino and fermion--sfermion, respectively. In Split
Supersymmetry, we will be mostly concerned with the case in which
decay into squarks, sleptons and one Higgs doublet are not
kinematically allowed, and thus $N_g=12$, $N_m=1$.

Since gravitinos in the early universe decay when
$\Gamma_{3/2}=H$, the temperature after decay is given by \beq
T_{3/2}=\left[ \frac{90~
\Gamma_{3/2}^2\mpl^2}{\pi^2g_*(T_{3/2})}\right]^{1/4} =\left[
\frac{10.75}{g_*(T_{3/2})}\right]^{1/4} \left(
\frac{m_{3/2}}{10^5\gev }\right)^{3/2}6.8 \mev , \label{gtemp}
\eeq where $g_*$ counts the effective number of degrees of
freedom.

%The gravitino number density $n_{3/2}$, in units of entropy density
%$s$,
%has been computed in supersymmetric QCD at finite temperature. By
%performing
%a hard thermal-loop resummation, where the contributions from soft
%momenta are regularized by considering an effective gluon propagator,
%the study in ref.~\cite{buch} has found
%\beq
%Y_{3/2}\equiv \frac{n_{3/2}}{s} = \frac{T_{R}}{10^{10}\gev}
%\left( 1+\frac{M_{\tilde g}^2}{12m_{3/2}^2}
%\right)
%9.4\times 10^{-13} .
%\label{gdens}
%\eeq
%Here $M_{\tilde g}$ is the zero-temperature gluino mass and
%$T_R$ is the reheating temperature after inflation (or, in general,
%the temperature after a significant entropy production). In \eq{gdens},
%we have assumed $T_R>\mtil$, so that all squarks participate
%in the gravitino production processes after reheating. The term
%proportional
%to the gluino mass becomes important when $M_*\ll \mpl$ and the
%Goldstino
%coupling is enhanced.

The gravitino number density in the early universe $n_{3/2}$,
before decay, evolves according to the Boltzmann
equation~\cite{giap} \beq \frac{d n_{3/2}}{dt} +3Hn_{3/2} =\left(
\gamma_{\rm sc} + \gamma_{\rm dec} \right) \left(
1-\frac{n_{3/2}}{n_{3/2}^{\rm eq}}\right). \label{bozzi} \eeq We
assume that SM and supersymmetric particles (other than the
gravitino) are in equilibrium. Two processes contribute to
\eq{bozzi}: gravitino emission in scatterings and decay of
thermalized supersymmetric particles into gravitinos.

The rate of gravitino production in scatterings $\gamma_{\rm sc}$
has been computed in supersymmetric QCD at finite temperature.
Performing a hard thermal-loop resummation, where the
contributions from soft momenta are regularized by considering an
effective gluon propagator, the study of ref.~\cite{buch} finds
\beq \gamma_{\rm sc} =\frac{T^6}{\mpl^2}C_{\rm sc} \eeq \beq
C_{\rm sc}=b\left( 1+\frac{b^2M_{\tilde g}^2}{12m_{3/2}^2}\right)
\left\{ 1+0.48n-0.74\left( 1+\frac{n}{3}\right) \ln \left[ b
\left( 1+\frac{n}{3}\right) \right] \right\} 8.4 \times 10^{-2}.
\label{frizzi} \eeq Here $n$ is the number of effective
quark-squark flavour multiplets ($n=6$ if squarks are light, and
$n=0$ in Split Supersymmetry with $T_R<\mtil $), and
$b(T)=2\alpha_s(T)/\alpha_s( M_{\tilde g})$ (in particular, $b(T)
\simeq 1$ for $T=10^{10}\gev$).
  The term proportional
to the gluino mass in \eq{frizzi} becomes important when $M_*\ll
\mpl$, since the Goldstino coupling is enhanced. Notice that we
have not included an analogous term proportional to the squark
masses. Indeed, while the effective Goldstino coupling to gluinos
is proportional to $M_{\tilde g}T/M_S^2$, the coupling to squarks
is proportional to $\mtil^2/M_S^2$~\cite{gosti}. Therefore, it
will give a contribution $C_{\rm sc}\propto \mtil^4
/(m_{3/2}T)^2$. Even for Split Supersymmetry, where $\mtil \gg
M_{\tilde g}$, this contribution can be neglected, as it is
subleading to the decay process we consider next.

Supersymmetric particles in thermal equilibrium can occasionally
decay into gravitinos, with a rate \beq \gamma_{\rm
dec}=\sum_{i=1}^N {\tilde n}^{\rm eq}_i \frac{K_1(z_i)}
  {K_2(z_i)} \Gamma_{\rm dec}^i,
\eeq where the sum extends over all supersymmetric particles in
the thermal bath with mass $\mtil_i$, and $z_i\equiv \mtil_i/T$.
The ratio of Bessel functions $K_1(z_i)/K_2(z_i)$ describes the
thermal average of the time dilatation factor $\mtil_i/E$. Using,
for simplicity, Maxwell-Boltzmann statistics for both scalars and
fermions, the equilibrium density of the supersymmetric particles
with 2 degrees of freedom is \beq {\tilde n}^{\rm eq}_i
=\frac{T^3}{\pi^2}z_i^2K_2(z_i). \eeq Finally, the decay width
into gravitinos is the same for both gauginos and sfermions \beq
\Gamma_{\rm dec}^i=\frac{\mtil_i^5}{48\pi m_{3/2}^2\mpl^2}. \eeq

It is convenient to rewrite \eq{bozzi} in terms of $Y_{3/2}\equiv
n_{3/2}/s$, the gravitino number density in units of entropy
density $s$. We find the differential equation \beq
\frac{dY_{3/2}}{dT}=-\frac{ \left( \gamma_{\rm sc} + \gamma_{\rm
dec} \right)}{HTs} \left( 1-\frac{Y_{3/2}}{Y_{3/2}^{\rm
eq}}\right) , \label{sprot} \eeq which can be easily integrated.
The term proportional to $\gamma_{\rm sc}$ is dominated by the
largest temperature $T_R$, to be interpreted as the reheat
temperature after inflation or, in general, as the temperature
after a significant entropy production. On the other hand, the
term proportional to $\gamma_{\rm dec}$ in \eq{sprot} is dominated
by temperatures $T\sim \mtil$. The solution is\footnote{We are
using $\int_0^\infty dz z^3 K_1(z)= 3\pi /2$.} \beq
Y_{3/2}=Y_{3/2}^{\rm eq}\left[ 1-\exp \left( -\frac{ Y_{3/2}^{\rm
sc} +Y_{3/2}^{\rm dec}}{Y_{3/2}^{\rm eq}}\right) \right] \simeq
{\rm min}\left(  Y_{3/2}^{\rm sc} +Y_{3/2}^{\rm dec}
~,~Y_{3/2}^{\rm eq} \right) \label{gravsd} \eeq \beq Y_{3/2}^{\rm
eq}=\frac{135 \zeta (3)}{4\pi^4g_*}\simeq 2\times 10^{-3} \eeq
\beq Y_{3/2}^{\rm sc}=\left. \frac{\gamma_{\rm
sc}}{Hs}\right|_{T=T_R} =\left( 1+\frac{M_{\tilde
g}^2}{12m_{3/2}^2}\right) \left( \frac{T_R}{10^{10}\gev}\right)
\left[ \frac{228.75}{g_*(T_R)}\right]^{3/2} 1.0\times 10^{-12}
\label{sprip0} \eeq \beq Y_{3/2}^{\rm dec}=\left.
\frac{135}{4\pi^3}\sum_{i=1}^N \frac{\Gamma_{\rm
dec}^i}{g_*H}\right|_{T=\mtil_i} = \left( \frac{\mtil}{\rm
TeV}\right)^3 \left( \frac{\rm GeV}{m_{3/2}}\right)^2
\left[\frac{228.75}{g_*(\mtil )}\right]^{3/2} \left(\frac{N}{46}
\right) 1.2 \times 10^{-13}. \label{sprip} \eeq The sum in
\eq{sprip} is over all supersymmetric particles such that
$m_{3/2}<\mtil_i <T_R$. If only the heavy states of Split
Supersymmetry satisfy this relation, then $N=46$.

In ordinary supersymmetry, the term $Y_{3/2}^{\rm dec}$ is
negligible, unless the gravitino is extremely light. In Split
Supersymmetry with large $\mtil$, the term  $Y_{3/2}^{\rm dec}$
can easily be the dominant source of gravitino production.

To proceed in our discussion, we consider different ranges of
gravitino masses.

\subsection{Case $m_{3/2}\gsim 10^5$ GeV}

This is the case in which we have to require $F_\phi \ll m_{3/2}$,
or else anomaly mediation gives an excessive contribution to
gaugino masses. This can be done whenever supersymmetry is
unbroken in the flat limit ($\mpl\to\infty$), as discussed in
sect.~\ref{anomed}.  In the range of $m_{3/2}$ considered in this
section, the gravitino is too heavy to play any role in collider
phenomenology. Also, gravitino decay occurs before the beginning
of nucleosynthesis, see \eq{gtemp}, and therefore it is not
harmful for the prediction of primordial-element abundances, in
contrast to the ordinary case. However, the gravitino can
significantly affect the neutralino relic abundance.

The lightest neutralino $\chi$ is stable and it decouples from the
thermal bath at a freeze-out temperature $T_f$ \beq
T_f=\frac{m_\chi}{x_f},~~~~x_f=28+\ln \left\{ \frac{\rm
TeV}{m_\chi} ~\frac{c}{10^{-2}} \left[
\frac{86.25}{g_*(T_f)}\right]^{1/2} \right\} . \label{sciaf1} \eeq
Here we have parametrized the non-relativistic $\chi$ annihilation
cross section as \beq \vev{\sigma_\chi v_{\rm rel}}
=\frac{c}{m_\chi^2}. \label{sciaf2} \eeq In Split Supersymmetry,
$c=3\times 10^{-3}$ for a mostly-higgsino $\chi$, and $c=1\times
10^{-2}$ for a mostly $W$-ino $\chi$ (including the effects of
coannihilation).

If $T_{3/2}>T_f$, all gravitino-decay products reach thermal and
chemical equilibrium. The $\chi$ relic abundance will be
unaffected by the gravitino decay process. We want to investigate
the opposite case, in which the $\chi$ produced in gravitino decay
are out of chemical equilibrium. This happens when \beq
T_{3/2}<T_f \Rightarrow ~~~~m_{3/2}< \left( \frac{m_\chi}{\rm
TeV}\right)^{2/3} \left[ \frac{g_*(T_{3/2})}{86.25}\right]^{1/6}
4.3\times 10^7\gev . \label{decdec} \eeq

The gravitino decay generates a population of $\chi$ with density
\beq Y_\chi (T_{3/2})\equiv \left.
\frac{n_\chi}{s}\right|_{T=T_{3/2}} =Y_{3/2} \left(
\frac{T_0}{T_{3/2}}\right)^3 , \label{chidd} \eeq where $Y_{3/2}$
is given in \eq{gravsd}. The factor $(T_0/T_{3/2})^3$, where $T_0$
is the temperature of the thermal bath before gravitino decay,
takes into account the entropy generated by the thermalized decay
products. We can compute $T_0$ by balancing the energy density
before and after the decay process, and thus by solving the
equation \beq \left( \frac{T_0}{T_{3/2}}\right)^3 \left(
\frac{T_0}{T_{3/2}} +\frac{4Y_{3/2}m_{3/2}}{3T_{3/2}}\right) =1.
\label{eqt0} \eeq For small $Y_{3/2}$, the solution is
$(T_0/T_{3/2})^3 \simeq 1-Y_{3/2}m_{3/2}/T_{3/2}$ and therefore
\beq Y_\chi (T_{3/2}) \simeq Y_{3/2} \left( 1-
Y_{3/2}\frac{m_{3/2}}{T_{3/2}} \right) ~~~~\hbox{(radiation
dominance)}. \eeq The $\chi$ number density is equal to the
gravitino number density before decay, and a small dilution factor
due to entropy production appears in $Y_\chi (T_{3/2})$.

For sufficiently large $T_R$ and low $T_{3/2}$ or for sufficiently
large $\mtil$ and small $m_{3/2}$, the universe eventually becomes
gravitino-dominated. This happens when \beq T_R\gsim \left(
\frac{m_{3/2}}{10^5\gev }\right)^{1/2} \left[
\frac{10.75}{g_*(T_{3/2})}\right]^{1/4} 5\times 10^{14}\gev
~~~~{\rm for}~Y_{3/2}^{\rm sc}>Y_{3/2}^{\rm dec}; \eeq \beq \mtil
\gsim \left( \frac{m_{3/2}}{10^5\gev }\right)^{5/6} \left[
\frac{10.75}{g_*(T_{3/2})}\right]^{1/12} 2\times 10^{8}\gev
~~~~{\rm for}~Y_{3/2}^{\rm dec}>Y_{3/2}^{\rm sc}. \eeq

Then the solution of \eq{eqt0} is $(T_0/T_{3/2})^3\simeq
3T_{3/2}/(4Y_{3/2} m_{3/2})$ and therefore \beq Y_\chi (T_{3/2})
\simeq \frac{3T_{3/2}}{4m_{3/2}} ~~~~\hbox{(gravitino dominance)}.
\label{grado} \eeq Indeed, we could have directly obtained this
result by noticing that, when the gravitino dominates the
universe, all radiation is produced by the decay process and
therefore $Y_\chi (T_{3/2}) =\rho_{\rm rad}/(s m_{3/2})$, which
coincides with \eq{grado}.

%Depending on whether the universe is gravitino-dominated before
%$\Gamma_{3/2}^{-1}$ or not, the $\chi$ density $Y_\chi \equiv n_\chi
%/s$
%from gravitino decay at $T=T_{3/2}$ is given by
%\beq
%Y_\chi (T_{3/2}) = \left\{ \begin{array}{ll}
%\frac{3T_{3/2}}{4m_{3/2}} & \hbox{gravitino dominance} \\
%Y_{3/2}                   & \hbox{radiation dominance}
%\end{array} \right.
%\label{chid}
%\eeq

Notice that, if the universe becomes gravitino-dominated, the
decay erases any primordial $\chi$ density. Region of parameter
space which are ruled out by an excessive dark-matter density, as
obtained by the standard thermal relic-density calculation, can
now be reconsidered. Even if the gravitino never dominates the
universe, its entropy production dilutes the primordial $Y_\chi$
by a factor $(T_0/T_{3/2})^3\simeq 1-Y_{3/2}m_{3/2}/T_{3/2}$.
Large entropy production leads to the potential problem of
diluting any cosmic baryon asymmetry produced at temperatures
larger than $T_{3/2}$. If the gravitino dominates the universe, we
necessarily have to invoke a low-temperature mechanism for
baryogenesis.

In balancing the energy density in \eq{eqt0} we have neglected the
$\chi$ contribution after gravitino decay. This is appropriate
because neutralinos, although out of chemical equilibrium, are in
kinetic equilibrium, since their rate for elastic scattering with
the thermal bath is larger than the Hubble rate
($T_{3/2}^3\sigma_{\rm el}>H$). This means that the $\chi$ will
rapidly lose their kinetic energy, becoming non-relativistic, with
an energy density $\rho_\chi =m_\chi n_\chi$. Since $n_\chi$ is
equal to $n_{3/2}$ before decay, $\rho_\chi$ is about $m_\chi
/m_{3/2}$ times smaller than the radiation energy density.

Since we are considering the case $T_{3/2}<T_f$, the particles
$\chi$ produced by the gravitino decay are out of chemical
equilibrium and satisfy the Boltzmann equation \beq
\frac{dn_\chi}{dt}+3Hn_\chi =-\vev{\sigma_\chi v_{\rm
rel}}n_\chi^2. \label{boltz} \eeq Using the variable
$z=T/T_{3/2}$, \eq{boltz} can be written as \beq
\frac{dY_\chi^{-1}}{dz}=-\beta^{-1}, \eeq \beq \beta \equiv \left.
\frac{H}{s\vev{\sigma_\chi v_{\rm rel}}} \right|_{T=T_{3/2}}
=\left(\frac{m_\chi}{\rm TeV}\right)^2 \left( \frac{10^5\gev
}{m_{3/2}}\right)^{3/2} \left( \frac{10^{-2}}{c}\right) \left[
\frac{86.25}{g_*(T_{3/2})}\right]^{1/4} 8.5 \times 10^{-10}. \eeq
The solution of this differential equation is \beq Y_\chi^{-1}
({\rm today})= Y_\chi^{-1}(T_{3/2}) +\beta^{-1}. \eeq This result
can be easily understood. If $n_\chi \vev{\sigma_\chi v_{\rm
rel}}<H$ at $T=T_{3/2}$ (\ie\ $Y_\chi (T_{3/2})<\beta$), then the
neutralinos can never annihilate, and their number density is just
diluted by expansion, and therefore $Y_\chi ({\rm today})= Y_\chi
(T_{3/2})$. On the other hand, if the opposite inequality holds,
the neutralinos will annihilate until their annihilation rate is
equal to the expansion rate $H$, and thus $Y_\chi ({\rm
today})=\beta$.

Using today's ratio of critical density versus entropy density
\beq {\left( \rho_c h^{-2}/s \right)_{\rm today}} =3.5\times
10^{-12}\tev , \label{rcsrat} \eeq we obtain \beq \Omega_\chi
h^2=\left. \frac{m_\chi Y_\chi }{\rho_ch^{-2}/s}\right|_{\rm
today} = \left(\frac{m_\chi}{\rm TeV}\right) \left[
Y_\chi^{-1}(T_{3/2}) +\beta^{-1} \right]^{-1} 2.8\times 10^{11},
\eeq where $Y_\chi^{-1}(T_{3/2})$ is given by \eq{chidd}.

\begin{figure}
\begin{center}
\epsfig{file=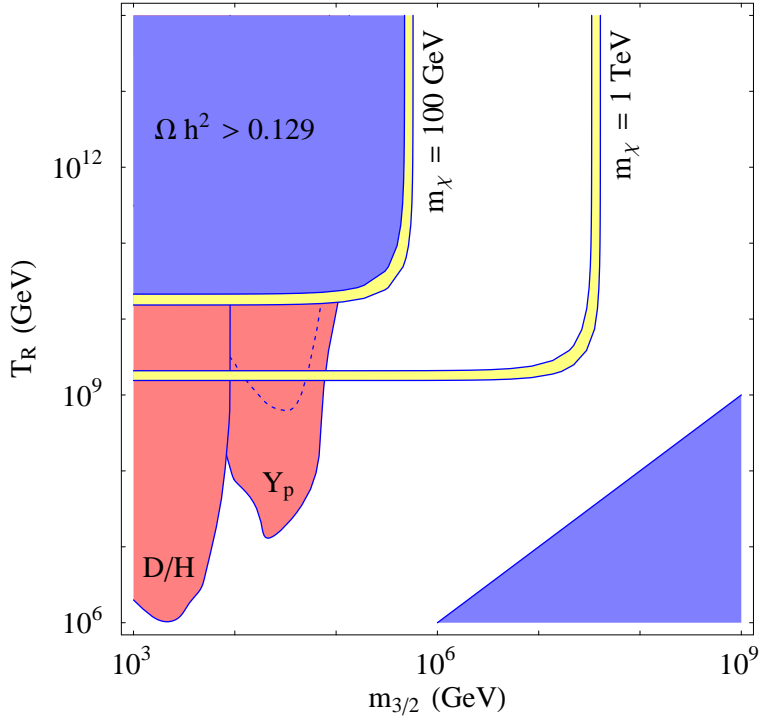,width=0.6\textwidth}
\end{center}
\caption{Case $\mtil >T_R$. The yellow bands correspond to a
neutralino density from gravitino decay $0.094<\Omega_\chi
h^2<0.129$, for $c=3\times 10^{-3}$, and $m_\chi=100\gev$ or
1~TeV. The region above the bands leads to an overdensity of dark
matter. The red region is excluded by the nucleosynthesis
constraints $D/H <3.6\times 10^{-5}$ and $Y_p<0.249$. The dashed
line corresponds to the weaker constraint $Y_p<0.253$. The region
in the bottom-right corner is excluded by the requirement
$T_R>m_{3/2}$.} \label{fig2}
\end{figure}

We first consider the case $\mtil > T_R$, in which $Y_{3/2}^{\rm
dec}$ can be neglected. In \fig{fig2} we show the parameter region
for which the non-thermal population of neutralinos from gravitino
decay can account for the required dark matter density in a
2-$\sigma$ range $0.094<\Omega_\chi h^2<0.129$~\cite{wmap}. The
result shown in \fig{fig2} is well explained by considering two
limiting cases. For $Y_\chi (T_{3/2})<\beta$, when $\chi$
annihilation is irrelevant, we find \beq \Omega_\chi h^2=
\left(\frac{m_\chi}{\rm TeV}\right)
\left(\frac{T_R}{10^{10}\gev}\right) 0.3 , \label{caso1} \eeq \beq
\Omega_\chi h^2 <0.129 \Rightarrow ~~~~T_R<
  \left(\frac{\rm TeV}{m_\chi}\right) 5\times 10^9\gev ,
\eeq which describes the horizontal branches of the curves in
\fig{fig2}. Notice that \eq{caso1} depends neither on the
microphysics $\chi$ interactions nor on the gravitino mass. For an
appropriate value of $T_R$, we can explain the correct dark-matter
density for neutralino masses which are inadequate to give the
right thermal abundance.

On the other hand, for $Y_\chi (T_{3/2})>\beta$
%(which is always the case
%when the gravitino dominates the universe before decaying).
we find \beq \Omega_\chi h^2=\left(\frac{m_\chi}{\rm TeV}\right)^3
\left( \frac{10^5\gev }{m_{3/2}}\right)^{3/2} \left(
\frac{10^{-2}}{c}\right) \left[
\frac{86.25}{g_*(T_{3/2})}\right]^{1/4} 2 \times 10^2,
\label{sciopmm} \eeq \beq \Omega_\chi h^2 <0.129 \Rightarrow ~~~~
m_{3/2}> \left(\frac{m_\chi}{\rm TeV}\right)^2 \left(
\frac{10^{-2}}{c}\right)^{2/3} \left[
\frac{86.25}{g_*(T_{3/2})}\right]^{1/6} 2 \times 10^7 \gev ,
\label{sciopm} \eeq which describes the vertical branches of the
curves in \fig{fig2}. This result depends on $m_{3/2}$ and on the
neutralino annihilation cross section, but it is independent on
$T_R$ and on the initial gravitino density (as long as it large
enough to satisfy $Y_\chi (T_{3/2})>\beta$). An appropriate $\chi$
relic density can be found for interesting values of $m_{3/2}$,
which are consistent with the requirement $T_{3/2}<T_{\rm dec}$,
see \eq{decdec}.

Next, we consider the mass range \beq T_R >\mtil >\left(
\frac{T_R}{10^{10}\gev}\right)^{1/3} \left(
\frac{m_{3/2}}{10^{5}\gev}\right)^{2/3} 4\times 10^6\gev ,
\label{condtr} \eeq in which $Y_{3/2}^{\rm dec}>Y_{3/2}^{\rm sc}$,
and therefore the gravitino density is independent of $T_R$.

\begin{figure}[!t]
\begin{center}
\epsfig{file=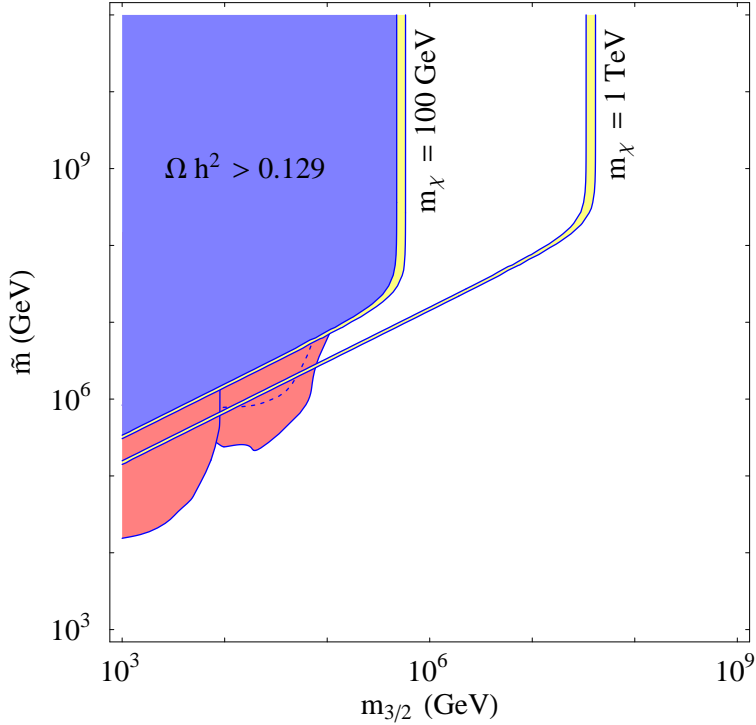,width=0.6\textwidth}
\end{center}
\caption{Case $T_R >\mtil >( {T_R}/10^{10}\gev )^{1/3} (
{m_{3/2}}/{10^{5}\gev})^{2/3} 4\times 10^6\gev$. The yellow bands
correspond to a neutralino density from gravitino decay
$0.094<\Omega_\chi h^2<0.129$, for $c=3\times 10^{-3}$, and
$m_\chi=100\gev$ or 1~TeV. The region above the bands leads to an
overdensity of dark matter. The red region is excluded by the
nucleosynthesis constraints $D/H <3.6\times 10^{-5}$ and
$Y_p<0.249$. The dashed line corresponds to the weaker constraint
$Y_p<0.253$.} \label{fig3}
\end{figure}

In this case, the parameter region in which neutralinos from
gravitino decay can account for the dark matter in shown in
\fig{fig3}. Again, the two branches of the curves can be simply
understood. The vertical branch is given by
eqs.~(\ref{sciopmm})--(\ref{sciopm}), while the oblique branch
corresponds to $Y_\chi ({\rm today})\simeq Y_{3/2}^{\rm dec}$ and
therefore \beq \Omega_\chi h^2=\left(\frac{m_\chi}{\rm TeV}\right)
\left(\frac{\mtil}{10^7\gev}\right)^3
\left(\frac{10^5\gev}{m_{3/2}}\right)^2 \left(\frac{N}{46}\right)
3 \eeq \beq \Omega_\chi h^2 <0.129 \Rightarrow ~~~~ \mtil <
\left(\frac{\rm TeV}{m_\chi}\right)^{1/3}
\left(\frac{m_{3/2}}{10^5\gev}\right)^{2/3}
\left(\frac{46}{N}\right)^{1/3} 3\times 10^6\gev . \eeq It is
interesting that, under the condition in \eq{condtr},
$\Omega_\chi$ from gravitino decay is completely determined in
terms of the mass parameters of the theory and it does not depend
on the cosmological initial condition parametrized by $T_R$.

Figures \ref{fig2} and \ref{fig3} illustrate how gravitino decay can
lead to the correct dark-matter neutralino density under conditions in
which the usual thermal abundance is insufficient.  This has important
implications for phenomenological analyses of Split Supersymmetry.
Indeed, we recall that the value of the dark-matter density plays a
crucial role in Split Supersymmetry, since it offers a rationale to
relate the gaugino--higgsino masses to the weak scale. Cases which
were excluded by consideration of thermal relic abundances (as an LSP
higgsino lighter than 1~TeV or an LSP W-ino lighter than
2~TeV~\cite{noi}) can be consistent with the observed amount of dark
matter, for appropriate values of $m_{3/2}$. 

\begin{figure}
\begin{center}
\epsfig{file=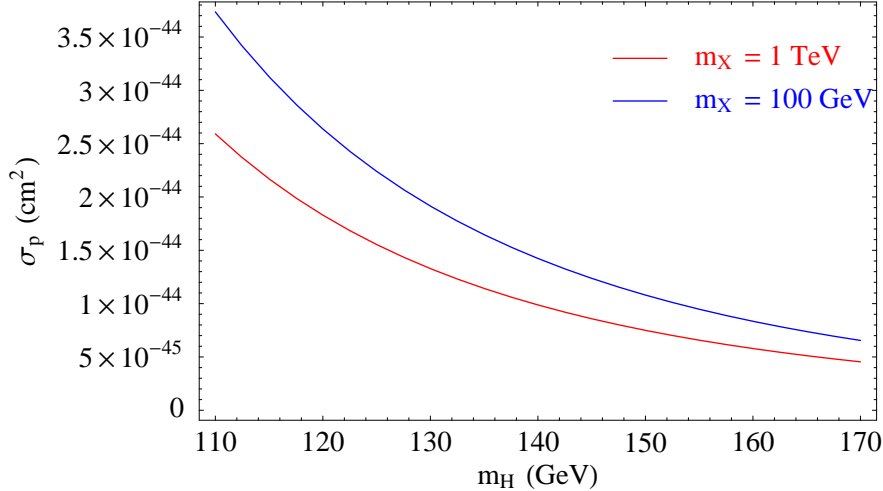,width=0.7\textwidth}
\end{center}
\caption{The maximum value of the spin-independent $\chi$ scattering cross
section off protons, as a function of the Higgs mass $m_H$ and for two
values of $m_\chi$. We have
assumed \eq{grel} at the chargino mass scale, and taken $\tan\beta = 10$.
No constraints on
$\Omega_\chi$ are used, assuming that gravitino decay accounts for
the correct value of $\Omega_\chi$.}
\label{det}
\end{figure}

The prospects for direct
and indirect dark-matter detection are also affected.  Neutralinos
with large annihilation cross sections can properly account for the
dark matter, because of gravitino decay. Since squarks are heavy, the
only contribution to spin-independent neutralino-nuclei interactions
comes from Higgs-boson exchange~\cite{Barbieri:1988zs}.  The $\chi$
scattering cross section off a proton is given by
% \begin{equation}
% \sigma_p =
% \left( N_{11}\tan\theta_W -N_{12}\right)^2 \left( N_{13}\cos\beta
%   -N_{14}\sin\beta \right)^2 \left( \frac{115~\gev} {m_H}\right)^4
% ~4\times 10^{-43}~{\rm cm}^2. \label{crossh} 
% \end{equation}
\begin{equation}
\sigma_p = \frac{8}{\pi}\left[ \frac{G_{\rm F} M_W m_p
    \mu_\chi}{9 m^2_H} \left( 2 +7\sum_{q=u,d,s}
    f^{(p)}_q \right) \gamma \right]^2 =  \left( \frac{115~\gev}
    {m_H}\right)^4 \gamma^2 ~5.4\times 10^{-43}\,{\rm cm}^2,
\label{crossh} 
\end{equation}
where~\cite{Gasser:1990ce} $f^{(p)}_u=0.023$, $f^{(p)}_d=0.034$,
$f^{(p)}_s=0.14$ and $\gamma$ measures the Higgs coupling with the LSP
neutralino
\begin{equation}
  \label{eq:mixing}
  \gamma = \frac{1}{g} \left( \tilde g_u N_{\chi 2}N_{\chi 4} -\tilde
  g_d N_{\chi 2}N_{\chi 3} -\tilde g'_u N_{\chi 1}N_{\chi 4} +\tilde
  g'_d N_{\chi 1}N_{\chi 3} \right) \,.
\end{equation}
Here $N_{\chi i}$ are the lightest neutralino components in standard
notations and ${\tilde g}_{u,d}$, ${\tilde g}_{u,d}^\prime$ are the
higgsino couplings (see sect.~\ref{edmsec}).
The coefficient $\gamma$ vanishes if $\chi$ is a pure Higgsino or
gaugino and in the limit $\mu, M_{1,2}\gg M_Z$ becomes (assuming
$M_2>M_1$)
\begin{equation}
  \label{eq:pert}
  \gamma = \cos\theta_W M_Z\frac{(\gtildpq+\gtilupq)M_1
  +2\gtilup\gtildp\mu}{g^2(\mu^2-M_1^2)} 
+{\cal O} \left( \frac{M_Z^2}{M_{1,2}^2},\frac{M_Z^2}{\mu^2}\right)
.
\end{equation}
The maximum value of $\gamma$ is reached when $M_1\simeq \mu$. In this
degenerate limit, eq.~(\ref{eq:pert}) is no longer valid, and it is
replaced by
\begin{equation}
  \label{eq:deg}
  \gamma = \frac{\gtilup+\gtildp}{2\sqrt{2}g} +\cos\theta_W\frac{
  M_Z}{8g^2}\left[ 2\frac{(\gtild+\gtilu)^2}{M_2-\mu} -
  \frac{(\gtildp-\gtilup)^2}{\mu} \right] 
+{\cal O} \left( \frac{M_Z^2}{M_{1,2}^2},\frac{M_Z^2}{\mu^2},
\frac{M_1-\mu}{\mu} \right) .
\end{equation}
The second term in the expansion is actually numerically important
because it is enhanced with respect to the leading term by a
coefficient $1/\tan\theta_W$. Notice that the maximal value of
$\gamma$, given by eq.~(\ref{eq:deg}), is actually achieved in a large
portion of the parameter space of Split Supersymmetry, leading to an
appropriate dark-matter thermal abundance~\cite{noi}. This is because
an efficient annihilation rate approximately requires $M_1\simeq\mu$.
In fig.~\ref{det} we show the spin-independent $\chi$ scattering cross
section off protons, without requiring any constraints on
$\Omega_\chi$ and therefore assuming that gravitino decay accounts for
the correct value of $\Omega_\chi$. The rate is within the reach of
future experiments, which can reach $10^{-44}$--$10^{-45}\,{\rm cm^2}$
for $m_\chi < 1\,{\rm TeV}$.

We also want to stress that the gravitino decay process does not
weaken the link between neutralino masses and the weak scale. This
link is based on the upper bound on the $\chi$ mass derived by the
requirement that the thermal relic abundance (for $s$-wave
annihilation) \beq \left( \Omega_\chi h^2 \right)_{\rm th} =\left(
   \frac{m_\chi}{\rm TeV} \right)^2 \left( \frac{ 10^{-2}}{c}\right)
\left( \frac{x_f}{28}\right) \left[
   \frac{86.25}{g_*(T_f)}\right]^{1/2} 3\times 10^{-2}
\label{omth} \eeq does not exceed the observed value. The
requirement $( \Omega_\chi h^2 )_{\rm th}<0.129$ gives \beq m_\chi
< \left( \frac{c}{ 10^{-2}}\right)^{1/2}
  \left( \frac{28}{x_f}\right)^{1/2}
\left[ \frac{g_*(T_f)}{86.25}\right]^{1/4}2\tev , \label{chilim}
\eeq where $c$ and $x_f$ are defined in
eqs.~(\ref{sciaf1})--(\ref{sciaf2}). The value of the non-thermal
$\Omega_\chi$ computed in this section has to be added to the
thermal result in \eq{omth} and therefore it can only lead to an
upper bound on $m_\chi$ which is {\it stronger} than \eq{chilim}.
Even in the case in which the gravitino dominates the universe and
dilutes the initial $\chi$ abundance, the upper bound on $m_\chi$
is tightened. Indeed, for a gravitino-dominated universe,
\eq{sciopmm} applies. Then we can interpret \eq{sciopm} as an
upper bound on $m_\chi$, as a function of $m_{3/2}$. This bound
becomes less stringent as $m_{3/2}$ grows, but a maximum allowed
value of $m_{3/2}$ is determined by the condition $T_{3/2}<T_f$ in
\eq{decdec}. For the value of $m_{3/2}$ corresponding to
$T_{3/2}=T_f$ we find an upper bound on $m_\chi$ which coincides
with \eq{chilim}, while for other values of $m_{3/2}$ the bound is
stronger. The only exception in which the neutralino mass could be
much larger than the value determined by \eq{chilim} occurs in the
extreme case when $T_R$ is of the order of $T_f$~\cite{rock}.

\subsection{Case $m_\chi < m_{3/2}\lsim 10^5$ GeV}

In this $m_{3/2}$ range, the anomaly-mediated contributions to soft
masses are acceptable and they can actually account for the entire
values of gaugino masses, since they give~\cite{anph} $M_1\simeq
m_{3/2}/100$, $M_2\simeq m_{3/2}/300$ and $M_{\tilde g} \simeq
m_{3/2}/40$. The collider phenomenology is then analogous to the one
of anomaly mediation, but with two very significant differences.
Squarks and sleptons can be much heavier than gauginos and higgsinos,
depending on the value of $M_*$. The parameter $\mu$ is not determined
by electroweak symmetry breaking to be larger than $M_2$ and therefore
the lightest supersymmetric particle could be a higgsino, or have a
higgsino component.

The gravitino lifetime $\tau_{3/2}$ is long enough to affect the
nucleosynthesis predictions. For the values of $\tau_{3/2}$
considered in this section, the strongest bounds come from
hadronic gravitino decays~\cite{stark,vari,mor}, while decays into
photons~\cite{vecchi,ellis} play no role. This is because,
injected photons have a high probability of scattering off
background electrons and photons, still very numerous at this
stage of nucleosynthesis, and they thermalize before having the
chance of photo-dissociating recently-created nuclei. But, since
we are considering the case $m_{3/2}>M_{\tilde g}$, we expect that
the gravitinos decay into hadrons with a branching fraction of
order unity, and therefore we turn to discuss the nucleosynthesis
limits from hadronic decay modes.

In the region $10\tev \lsim m_{3/2} \lsim 100\tev$ ($30~ {\rm sec}
\gsim \tau_{3/2} \gsim 3\times 10^{-2}~{\rm sec}$), quarks and
gluons produced by gravitino decay quickly hadronize, and the
hadrons lose their kinetic energy by scattering off background
electrons and photons without causing hadro-dissociation of
nuclei. However, there can still be an effect on nucleosynthesis,
because the slowed-down hadrons interact with ambient nucleons,
converting neutrons into protons and viceversa. This will change
the ratio $n/p$, which was fixed after weak interactions have
frozen out. Since protons are more abundant than neutrons, the
gravitino decay products will increase the ratio $n/p$ and
therefore the predicted $^4$He fraction. Using the results
presented in ref.~\cite{mor}, we set the bounds on $T_R$ in
\fig{fig2} and on $\mtil$ in \fig{fig3} which correspond to a 95\%
CL limit on the $^4$He fraction $Y_p<0.249$~\cite{olive}. To
illustrate the strong sensitivity of these bounds on $Y_p$, we
also show in figs.~\ref{fig2} and \ref{fig3} the weaker limit
corresponding to $Y_p<0.253$, obtained from the study in
ref.~\cite{rus}.

In the region $1\tev \lsim m_{3/2} \lsim 10\tev$ ($3000~ {\rm sec}
\gsim \tau_{3/2} \gsim 30~{\rm sec}$), injected hadrons can
directly destroy nuclei. In particular, hadro-dissociation of
$^4$He leads to an overabundance of deuterium. Using a 95\% CL
limit $D/H <3.6\times 10^{-5}$~\cite{deut} and the calculation of
ref.~\cite{mor}, we obtain the bounds shown in figs.~\ref{fig2}
and \ref{fig3}. Notice that in our case photo-dissociation of
$^4$He is irrelevant, since it starts to be effective only after
about $10^5$ seconds~\cite{ellis}. Overproduction of $^6$Li by
hadro-dissociation starts to become competitive with the bounds
from deuterium only for $m_{3/2}$ close to 1 TeV.

 From the results shown in figs.~\ref{fig2} and \ref{fig3},
it is apparent that nucleosynthesis provides strong constraints on
$T_R$ and $\mtil$. In particular, from this bound we infer that,
in the range of $m_{3/2}$ considered in this section, the
gravitino decay process cannot generate a neutralino density
sufficient to explain the dark matter. In this case, the
appropriate value of $\Omega_\chi h^2$ has to be generated by the
neutralino thermal relic abundance.

\subsection{Case $m_{3/2}<m_\chi$}
\label{lightgravitino}

In this mass region, the gravitino is the LSP. In the absence of
$R$-parity breaking, it is stable and its contribution to the
present energy density is \beq \Omega_{3/2}h^2 =\frac{m_{3/2}
Y_{3/2}}{\left( \rho_c h^{-2}/s \right)_{\rm today}}, \eeq where
$Y_{3/2}$ is given in \eq{gravsd} and today's ratio of critical
density to entropy is given in \eq{rcsrat}. The abundance
$Y_{3/2}$ has three distinct behaviors. If $T_R$ or $\mtil$ are
large enough, gravitinos go in equilibrium and \beq \Omega_{3/2}
h^2 = \left( \frac{m_{3/2}}{\rm keV}\right) \left(
\frac{228.75}{g_*}\right) 0.5 ~~~{\rm for}~Y_{3/2}=Y_{3/2}^{\rm
eq}, \eeq \beq \Omega_{3/2} h^2 <0.129 \Rightarrow ~~~~
m_{3/2}<\left( \frac{g_*}{228.75}\right) 0.3~{\rm keV} .
\label{omg0} \eeq Here $g_*$ has to be computed at the gravitino
decoupling temperature. If \eq{condtr} holds, than the decay
processes dominate the gravitino density and \beq \Omega_{3/2}
h^2= \left( \frac{\mtil}{\rm TeV}\right)^3 \left( \frac{\rm
MeV}{m_{3/2}}\right) \left[ \frac{228.75}{g_*(\mtil)}\right]^{3/2}
\left( \frac{N}{46}\right) 3\times 10^{-2} ~~~{\rm
for}~Y_{3/2}=Y_{3/2}^{\rm dec}, \label{caz1} \eeq \beq
\Omega_{3/2} h^2 <0.129 \Rightarrow ~~~~ \mtil < \left(
\frac{m_{3/2}}{\rm MeV}\right)^{1/3} \left[
\frac{g_*(\mtil)}{228.75}\right]^{1/2} \left(
\frac{46}{N}\right)^{1/3} 2\tev . \label{omg1} \eeq Otherwise, the
scattering processes dominate the gravitino production and \beq
\Omega_{3/2} h^2= \left( \frac{\rm MeV}{m_{3/2}}\right) \left(
\frac{M_{\tilde g}}{\rm TeV}\right)^2 \left(
\frac{T_R}{10^{10}\gev}\right) \left[
\frac{228.75}{g_*(T_R)}\right]^{3/2} 2\times 10^4 ~~~{\rm
for}~Y_{3/2}=Y_{3/2}^{\rm sc} , \label{caz2} \eeq \beq
\Omega_{3/2} h^2 <0.129 \Rightarrow ~~~~ T_R < \left(
\frac{m_{3/2}}{\rm MeV}\right) \left( \frac{\rm TeV}{M_{\tilde
g}}\right)^2 \left[ \frac{g_*(T_R)}{228.75}\right]^{3/2}
  5\times 10^4\gev .
\label{omg2} \eeq Finally the decay of the next-to-lightest
supersymmetric particle (NLSP) provides an additional source of
relic gravitinos, \beq \Omega_{3/2} h^2 =\frac{m_{3/2}}{m_\chi}
\left( \Omega_\chi h^2 \right)_{\rm th} ~~~{\rm for~NLSP~decay},
\label{ompif} \eeq \beq \Omega_{3/2} h^2 <0.129 \Rightarrow ~~~~
m_{3/2}< \left( \frac{\rm TeV}{m_\chi}\right) \left(
\frac{c}{10^{-2}}\right) \left( \frac{28}{x_f}\right) \left[
\frac{g_*(T_f)}{86.25}\right]^{1/2} 4\times 10^3\gev . \eeq We
have used the thermal abundance of NLSP $( \Omega_\chi h^2 )_{\rm
th}$ given by \eq{omth}. Around the inflationary epoch, other
sources of non-thermal gravitino production can be
present~\cite{nonth}, like the effect of the classical
gravitational background on the vacuum state during the evolution
of the universe. We will ignore here these mechanisms, which are
more model-dependent.

The limits from eqs.~(\ref{omg1}) and (\ref{omg2}) become very
stringent for small gravitino masses. Of course, whenever the
inequalities are nearly saturated, then gravitinos can account for
the dark matter. However if gravitinos, rather than neutralinos,
form the dark matter, then we lose the connection between the
higgsino-gaugino masses and the weak scale, which is a critical
ingredient of Split Supersymmetry.

Actually an upper bound on $m_\chi$ persists, even when the
gravitino is the dark matter particle. Indeed, $\Omega_{3/2}$ from
NLSP decays scales like $\Omega_{3/2}\propto m_{3/2} m_\chi$, see
\eq{ompif}. Large values of $m_\chi$ require small values of
$m_{3/2}$ to avoid an excessive gravitino dark-matter density. On
the other hand, the contribution to the gravitino energy density
from thermal processes scales like $\Omega_{3/2} \propto
m_{3/2}^{-1}$, see eqs.~(\ref{caz1}) and (\ref{caz2}), and a small
$m_{3/2}$ gives an excessive $\Omega_{3/2}$. Depending on the
value of $T_R$ and $\mtil$, the upper bound on $m_\chi$ can be
very restrictive. However, it is not sufficient to unescapably tie
the gaugino-higgsino masses to the weak scale. For instance, in
the extreme case $\mtil >T_R\sim M_{\tilde g}\sim m_\chi$, a
dark-matter gravitino of about 10~GeV is compatible with $m_\chi$
as large as 100~TeV.

The NLSP lifetime \beq \tau_\chi
=\frac{48\pi\mpl^2m_{3/2}^2}{m_\chi^5}= \left( \frac{m_{3/2}}{\rm
MeV}\right)^2 \left( \frac{\rm TeV}{m_\chi}\right)^5 6\times
10^{-7} ~{\rm sec} \eeq has a strong dependence on $m_\chi$ and
$m_{3/2}$, but the decay can occur at late times and affect
nucleosynthesis predictions and the cosmic microwave background.
Large ranges of parameters are excluded~\cite{mor,ellis,nuk}. In
particular, the case of a weak-scale gravitino is ruled out as a
dark-matter candidate for a neutralino NLSP~\cite{gradm}.

\subsection{Collider Phenomenology for Gravitino LSP}

%When the gravitino is the LSP, the contribution to the mass spectrum
%from anomaly-mediation can be safely neglected.
The collider phenomenology with a gravitino LSP can be quite different
than in ordinary low-energy supersymmetry, if the NLSP decay occurs
within measurable distances. The NLSP can decay into $\gamma$, $Z$, or
$H$ and missing energy. The signals will be similar to those of gauge
mediation~\cite{gaugemed}, in the limit of heavy scalars.

A peculiarity of Split Supersymmetry, in the light-gravitino
scenario, is represented by the gluino. Although the gluino is not
the LSP, its dominant decay mode could be into gravitinos. Indeed,
the decay width into gravitinos $\tilde G$ is \beq \Gamma \left(
{\tilde g} \to g {\tilde G} \right) = \frac{M_{\tilde
g}^5}{16\pi\mpl^2m_{3/2}^2} \left( 1-\frac{m_{3/2}^2} {M_{\tilde
g}^2}\right)^3 \left(
  1+3\frac{m_{3/2}^2}
{M_{\tilde g}^2}\right) . \eeq The ordinary gluino decay is
suppressed by the large value of $\mtil$ and by the 3-body phase
space, \beq \Gamma \left( {\tilde g} \to q {\bar q} \chi \right) =
\frac{\alpha \alpha_s {\cal N}M_{\tilde g}^5}{192\pi\sin^2\theta_W
\mtil^4} . \eeq Here $\cal N$ is a factor which appropriately
counts all decay channels into different quarks and charginos or
neutralinos. For $m_{3/2}\simeq \mtil^2/\mpl$ and
$m_{3/2}<M_{\tilde g}$, the dominant decay mode is ${\tilde g} \to
g {\tilde G}$, and each gluino leads to a single jet with
characteristic energy distribution. The same final state could be
mimicked, in a theory with heavy gravitino, by the radiative
process ${\tilde g} \to g \chi$, where $\chi$ is the neutralino
LSP. Such process is induced by the one-loop generated operator
${\bar \chi} \sigma^{\mu \nu} \gamma_5 {\tilde g}^a G_{\mu
\nu}^a$, where $\gamma_5$ arises because of the Majorana nature of
${\tilde g}$ and $\chi$. This operator is C (and P) odd, and
therefore it is not generated when left and right squarks are mass
degenerate. Therefore, we expect a certain suppression. When the
decay mode ${\tilde g} \to g {\tilde G}$ dominates over the
tree-level 3-body process, it is also dominant with respect to the
one-loop 2-body decay. A dominant gluino decay into a single jet
and missing energy is a very distinctive signature of Split
Supersymmetry with direct mediation.

\section{CP Violation and Electric Dipole Moments}
\label{edmsec}

Split Supersymmetry resolves the difficulties with flavour and CP
violation, encountered in generic supersymmetric models, because
squarks and sleptons are taken to be very heavy, with masses of
the order of $\mtil$. Nevertheless, the effective theory below
$\mtil$ still contains physical CP-violating phases.

Consider the interaction Lagrangian of higgsinos (${\tilde
H}_{u,d}$), W-ino ($\tilde W$) and B-ino ($\tilde B$) with the
Higgs doublet $H$
 \bea
-{\cal L}&=& \frac{M_2}{2} {\tilde W}^a {\tilde W}^a
+\frac{M_1}{2} {\tilde B} {\tilde B}
+\mu {\tilde H}_u^T\epsilon {\tilde H}_d \nonumber \\
&+&\frac{H^\dagger}{\sqrt{2}}\left( \gtilu \sigma^a {\tilde W}^a
+\gtilup {\tilde B} \right) {\tilde H}_u
+\frac{H^T\epsilon}{\sqrt{2}}\left( -\gtild \sigma^a {\tilde W}^a
+\gtildp {\tilde B} \right) {\tilde H}_d +{\rm h.c.} ,
\label{lagr} \eea where $\epsilon =i\sigma_2$ and, at the scale
$\mtil$, \beq \gtilu^{(\prime )}=g^{(\prime )}\sin\beta ,~~~~
\gtild^{(\prime )}=g^{(\prime )}\cos\beta . \label{grel} \eeq The
gaugino and higgsino masses $M_{1,2}$ and $\mu$ are, in general,
complex parameters. They can all be made real by a field
redefinition. However, the phase redefinition of gauginos and
higgsinos will induce two {\it independent} phases in the
couplings $\gtilu$ and $\gtild$ (and analogously in ${\tilde
g}_{u,d}^\prime$). There is still the freedom to transform $H\to
e^{i\alpha}H$, ${\tilde H}_u \to e^{i\beta}{\tilde H}_u$, ${\tilde
H}_d \to e^{-i\beta}{\tilde H}_d$, without reintroducing phases in
$M_{1,2}$ and $\mu$. However, this will induce the transformations
$\gtilu \to e^{i(\beta -\alpha )}\gtilu$, $\gtild \to e^{i(\alpha
-\beta )}\gtild$, and therefore only one of the two independent
phases can be eliminated. In conclusion, the two quantities ${\rm
Im}(\gtilu^* \gtild^* M_2 \mu)$ and ${\rm Im}(\gtilu^{\prime *}
\gtild^{\prime *} M_1 \mu)$ are invariant under any field-phase
redefinition and therefore correspond to physical CP-violating
effects\footnote{If the heavy Higgs fields integrated out at the
scale $\mtil$ violate CP, the couplings $\gtilu^{(\prime )}$ and
$\gtild^{(\prime )}$ can in principle have independent phases. In
this case, there is a third CP-violating invariant, given by ${\rm
Im}(\gtilu \gtilu^{\prime *} \gtild^* \gtildp)$. However, this
invariant vanishes, if the CP violation appears only in the
parameter $\beta$ defined in \eq{grel} since, in this case, ${\rm
arg}(\gtilu )= {\rm arg}(\gtilup )$ and  ${\rm arg}(\gtild )= {\rm
arg}(\gtildp )$.}. In most models of supersymmetry breaking, the
phases of the different gaugino masses $M_{1,2}$ are equal, and
therefore there is actually a single CP-violating invariant.
Notice also that the invariants vanish in the limit in which
$M_{1,2}$ and $\mu$ have opposite phases (in the notation of
\eq{lagr} above) or when either $\gtilu$ or $\gtild$ are absent
(as it is approximately true in the limit of large $\tan\beta$).

The new phases are in the chargino--neutralino sector which does
not couple at tree level to quarks and leptons, and therefore CP
violation in processes involving SM fermions occurs only at two
loops. The most interesting effect appears in the fermion electric
dipole moments (EDM). These are induced by the effective operator
$\bar f \sigma_{\mu \nu} \gamma_5 f F^{\mu \nu}$, where $f$ is a
generic fermion ($f=e$ for the electron EDM, and $f=u,d,s$ for the
neutron EDM). As gluinos carry no phases, there are no
contributions from either chromomagnetic or 3-gluon
operators~\cite{weinb} at the leading order.

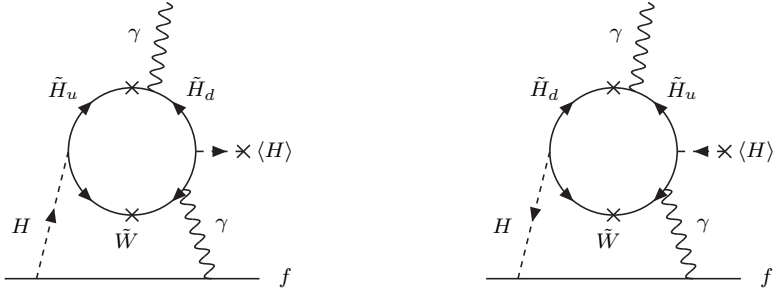
\begin{figure}
\begin{center}
\begin{picture}(160,140)(-80,-60)
\SetScale{1.2} \ArrowArcn(0,0)(20,180,90)
\ArrowArcn(0,0)(20,0,-90) \ArrowArc(0,0)(20,0,90)
\ArrowArc(0,0)(20,180,270) \Line(-2,18)(2,22)\Line(2,18)(-2,22)
\Line(-2,-18)(2,-22)\Line(2,-18)(-2,-22) \Line(-40,-40)(40,-40)
\DashArrowLine(-30,-40)(-20,0){2} \DashArrowLine(20,0)(35,0){2}
\Line(33,-2)(37,2)\Line(37,-2)(33,2) \Photon(25,-40)(16,-12){2}{6}
\Photon(6.8404,18.7939)(10.8404,46.7939){2}{6}
\Text(-26,24)[]{\scriptsize$\tilde H_u$}
\Text(26,24)[]{\scriptsize$\tilde H_d$}
\Text(54,0)[]{\scriptsize$\left\langle H \right\rangle$}
\Text(-42,-28)[]{\scriptsize$H$}
\Text(34,-28)[]{\scriptsize$\gamma$}
\Text(-2,-33)[]{\scriptsize$\tilde W$}
\Text(1,44)[]{\scriptsize$\gamma$} \Text(58,-48)[]{\scriptsize$f$}
\end{picture}
\hspace{5mm}
\begin{picture}(160,140)(-80,-60)
\SetScale{1.2} \ArrowArcn(0,0)(20,180,90)
\ArrowArcn(0,0)(20,0,-90) \ArrowArc(0,0)(20,0,90)
\ArrowArc(0,0)(20,180,270) \Line(-2,18)(2,22)\Line(2,18)(-2,22)
\Line(-2,-18)(2,-22)\Line(2,-18)(-2,-22) \Line(-40,-40)(40,-40)
\DashArrowLine(-20,0)(-30,-40){2} \DashArrowLine(35,0)(20,0){2}
\Line(33,-2)(37,2)\Line(37,-2)(33,2) \Photon(25,-40)(16,-12){2}{6}
\Photon(6.8404,18.7939)(10.8404,46.7939){2}{6}
\Text(-26,24)[]{\scriptsize$\tilde H_d$}
\Text(26,24)[]{\scriptsize$\tilde H_u$}
\Text(54,0)[]{\scriptsize$\left\langle H \right\rangle$}
\Text(-42,-28)[]{\scriptsize$H$}
\Text(34,-28)[]{\scriptsize$\gamma$}
\Text(-2,-33)[]{\scriptsize$\tilde W$}
\Text(1,44)[]{\scriptsize$\gamma$} \Text(58,-48)[]{\scriptsize$f$}
\end{picture}
\end{center}
\caption{Feynman diagrams contributing to the fermion EDM in Split
Supersymmetry. To better illustrate the structure of the
interactions, we consider current eigenstates with insertions of
$M_2$, $\mu$, and $\vev{H}$ denoted by crosses. Two other diagrams
with reversed directions of chargino arrows are not shown.}
\label{figfey}
\end{figure}

The fermion EDM is generated by the two-loop diagrams shown in
\fig{figfey}. Other diagrams can also be constructed, by replacing
the $\gamma$ and $H$ internal lines with various combinations of
Higgs and weak gauge bosons~\cite{edm2l}. However, the diagrams
shown in \fig{figfey} form a gauge-invariant set which is expected
to give the leading contribution. In particular, notice that
diagrams with $Z$ bosons give a contribution to the electron EDM
which is suppressed by the $Z$ vector coupling proportional to
$1-4\sin^2\theta_W$. The contribution of the diagrams in
\fig{figfey} to the EDM $d_f$ of a fermion with electric charge
$Q_f$ and mass $m_f$ is \beq \frac{d_f}{e} =\frac{\alpha Q_f m_f
gK_{\rm QED}}{32\sqrt{2}\pi^3M_Wm_H^2} ~{\rm Im}~\sum_{i=1}^2
\left( \gtild^*  U_{i2} V_{i1} + \gtilu^*  U_{i1} V_{i2} \right)
m_{\chi^+_i} f\left( \frac{m_H^2}{ m_{\chi^+_i}^2}\right) \eeq
\bea f(x)&=&\frac{2\sqrt{x}}{\sqrt{x-4}}\left[ \ln x \ln
\frac{\sqrt{x-4}+\sqrt{x}}{\sqrt{x-4}-\sqrt{x}} +{\rm Li}_2\left(
\frac{2\sqrt{x}}{\sqrt{x}-\sqrt{x-4}}\right) -{\rm Li}_2\left(
\frac{2\sqrt{x}}{\sqrt{x}+\sqrt{x-4}}\right)
\right] \nonumber \\
&=& \left( 2-\ln x\right) x +\left( \frac{5}{3}-\ln x\right)
\frac{x^2}{6} +\order{x^3}. \eea Here $K_{\rm QED}$ is the
leading-logarithm QED correction in the running from the scale of
the heavy particles to $m_f$ (or $m_n$ for the neutron
EDM)~\cite{edmqed} \beq K_{\rm QED}=1-\frac{4\alpha}{\pi}\ln
\frac{m_H}{m_f}. \eeq We work in a general basis in which
$\gtilu$, $\gtild$, $M_2$, and $\mu$ are all complex. The matrices
$U$ and $V$ are defined such that $U^*{\cal M}_{\chi^+} V^\dagger$
is diagonal with real and positive entries, where ${\cal
M}_{\chi^+}$ is the chargino mass matrix \beq {\cal M}_{\chi^+}
=\pmatrix{M_2 & \sqrt{2} M_W \gtilu /g \cr \sqrt{2} M_W \gtild /g
& \mu }. \label{mcha} \eeq

We can explicitly write the matrices $U$ and $V$ as \beq
U=\pmatrix{c_R e^{i\phi_1} & s_R e^{i(\phi_1 -\delta_R)}\cr
-s_Re^{i\phi_2} & c_R e^{i(\phi_2 -\delta_R)}} ~~~~ V=\pmatrix{c_L
& s_L e^{-i\delta_L}\cr -s_L & c_L e^{-i\delta_L}} \eeq \beq
%\tan 2\theta_R =\frac{2|X_R|}{1+|X_L|^2-|X_R|^2} ,~~~~
\tan 2\theta_{L,R} =\frac{2|X_{L,R}|}{1+|X_{R,L}|^2-|X_{L,R}|^2}
,~~~~ e^{i\delta_{L,R}}=\frac{X_{L,R}}{|X_{L,R}|},
%\delta_R ={\rm arg}X_R,~~~~
%\delta_L ={\rm arg}X_L,
\eeq \beq X_L=\frac{\sqrt{2} M_W (\gtilu^* M_2 +\gtild
\mu^*)}{g(|M_2|^2-|\mu |^2)}, ~~~~ X_R=\frac{\sqrt{2} M_W
(\gtild^* M_2 +\gtilu \mu^*)}{g(|M_2|^2-|\mu |^2)}, \eeq where
$s_{L,R}\equiv \sin\theta_{L,R}$ and $c_{L,R}\equiv
\cos\theta_{L,R}$. The phases $\phi_1$ and $\phi_2$ are chosen
such that $m_{\chi^+_i}$ are real and positive. Using the
diagonalization properties of the matrices $U$ and $V$, we obtain
our final expression for the fermion EDM: \beq \frac{d_f}{e}
=\frac{\alpha Q_f m_f gK_{\rm QED}}{32\sqrt{2}\pi^3M_Wm_H^2} ~{\rm
Im}~\left( \gtild^* c_Ls_Re^{-i\delta_R} +\gtilu^*
c_Rs_Le^{-i\delta_L}\right) e^{i\phi_1} m_{\chi_1^+} \left[
f\left( \frac{m_H^2}{ m_{\chi^+_1}^2}\right) - f\left(
\frac{m_H^2}{ m_{\chi^+_2}^2}\right) \right]  . \label{eddm} \eeq
Notice that $d_f$ vanishes if $m_{\chi^+_1}=m_{\chi^+_2}$. Indeed,
if the two mass eigenvalues are equal, than $|X_L|=|X_R|=0$, and
this implies that the CP invariant ${\rm Im}(\gtilu^* \gtild^* M_2
\mu )$ vanishes.

To have a better understanding of this result, we expand \eq{eddm}
in small $X_{L,R}$. This expansion is valid for chargino masses
larger than $M_W$, and not degenerate ($|M_2-\mu|>M_W$). In this
case, we have $s_{L,R}\simeq |X_{L,R}|$, $\exp (i\phi_1)\simeq
M_2/|M_2|$, $m_{\chi_1^+}=|M_2|$ and $m_{\chi_2^+}=|\mu |$. Thus
\beq \frac{d_f}{e} =\frac{\alpha Q_f m_f K_{\rm QED} {\rm
Im}(\gtilu^* \gtild^* M_2 \mu )}
{16\pi^3m_H^2(|M_2|^2-|\mu|^2)}\left[ f\left( \frac{m_H^2}{
|M_2|^2}\right) - f\left( \frac{m_H^2}{ |\mu|^2}\right) \right]
+{\cal O}\left( \frac{M_W^2}{|M_2\mu|} \right) \label{df43} \eeq
Notice that \eq{df43} is exactly proportional to the CP-violating
invariant. We can further simplify the result by considering the
limit $|M_2|,|\mu|> m_H$: \beq \frac{d_f}{e} =-\frac{\alpha Q_f
m_f K_{\rm QED} \gtilu \gtild \sin\Phi} {16\pi^3M_2\mu} \left( \ln
\frac{M_2\mu}{m_H^2}+2+\frac{M_2^2+\mu^2}{M_2^2-\mu^2}\ln
\frac{\mu}{M_2}\right) +{\cal O}\left( \frac{M_W^2}{M_2\mu},
\frac{m_H^2}{M_2\mu} \right) . \label{defin} \eeq Here (and in the
following) all variables indicate the absolute value of the
corresponding quantity and $\Phi ={\rm arg}(\gtilu^* \gtild^* M_2
\mu )$.

The result in \eq{defin} can be rederived with a simple argument.  By
making a chiral rotation of the chargino fields we can eliminate all
phases in the mass matrix in \eq{mcha}. Since this transformation is
anomalous, it generates a term in the Lagrangian
\beq 
\frac{e^2}{16\pi^2}{\rm arg}\left( {\rm det} {\cal
M}_{\chi^+} \right) F_{\mu \nu} {\tilde F}^{\mu\nu} . 
\eeq 
We can expand ${\rm arg}( {\rm det} {\cal M}_{\chi^+} )$ in powers of
the Higgs background field using $M_W\to g|H|/\sqrt{2}$ in \eq{mcha},
\beq 
{\rm arg}\left[ {\rm det} {\cal M}_{\chi^+} (H)\right] = {\rm
  arg}(M_2\mu) + \frac{\gtilu \gtild \sin\Phi}{M_2\mu} H^\dagger H
+\order{H^4} . 
\eeq 
Therefore, the chiral rotation has induced the operator
\beq 
{\cal O}_H =\frac{c_H}{\Lambda^2}H^\dagger HF_{\mu \nu}
{\tilde F}^{\mu\nu}, \label{opcaz} 
\eeq 
\beq
c_H=\frac{\alpha}{4\pi}\gtilu\gtild \sin\Phi ,~~~~\Lambda^2=M_2\mu .
\eeq 
Through QED renormalization, the operator ${\cal O}_H$ mixes with the
EDM operator~\cite{mix}, generating 
\beq 
\frac{d_f}{e} =-\frac{Q_f
  m_fc_H} {4\pi^2\Lambda^2}\ln \frac{\Lambda^2}{m_H^2}. 
\eeq 
This correctly reproduces the leading-logarithm behaviour of
\eq{defin}.

\begin{figure}
\begin{center}
\epsfig{file=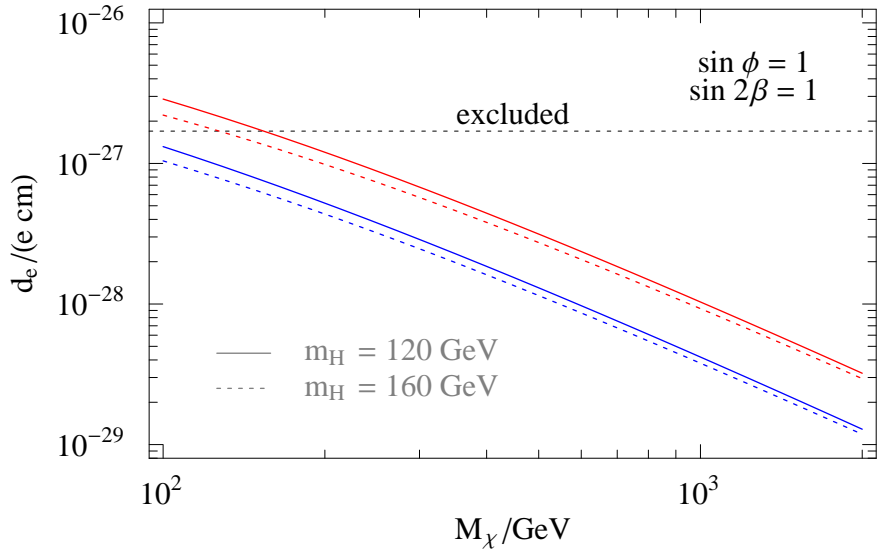,width=0.7\textwidth}
\end{center}
\caption{The prediction of the electron EDM in Split
Supersymmetry. We plot $d_e /(\sin\Phi \sin 2\beta)$ as a function
of the lightest chargino mass $m_{\chi^+_1}$. The CP-violating
phase is $\Phi ={\rm arg}(\gtilu^*\gtild^* M_2\mu)$ and we have
used \eq{grel} at the chargino mass scale. Solid lines correspond
to $m_H=120\gev$ and dashed lines to $m_H=160\gev$. The top two
lines correspond to $m_{\chi^+_2}/ m_{\chi^+_1}=1.5$ and the
bottom two lines to $m_{\chi^+_2}/ m_{\chi^+_1}=4$. The horizontal
line shows the present limit $d_e < 1.7\times 10^{-27}~ e$~cm at
95\% CL~\cite{edmex}.} \label{fig5}
\end{figure}

The prediction for the electron EDM in Split Supersymmetry is
shown in \fig{fig5}, taking the relations in \eq{grel} to be
approximately valid at the chargino mass scale. The deviation from
a straight-line behaviour of the curves in \fig{fig5} is a result
of the logarithmic enhancement explained above. For weak-scale
chargino masses and a maximal CP-violating phase, the result is
very close to the present experimental limit $d_e < 1.7\times
10^{-27}~ e$~cm at 95\% CL~\cite{edmex}. In ordinary low-energy
supersymmetry, EDMs are generated at one loop and therefore small
phases $(\lsim 10^{-2})$ are necessary to reconcile theory with
experiments. Because of the two-loop suppression, Split
Supersymmetry makes the exciting prediction that EDMs are on the
verge of being experimentally tested, if phases take their most
natural value of order unity.

EDM experiments are therefore at the frontier of testing Split
Supersymmetry. They may reveal hints of new physics even before
the start of the LHC. Ongoing and next generation experiments plan
to improve the EDM sensitivity by several orders of magnitude
within a few years. For example, DeMille and his Yale
group~\cite{demille} will use the molecule PbO to improve the
sensitivity of the electron EDM  to $10^{-29}~e$~cm within three
years, and possibly to $10^{-31}~e$~cm within five years.
Lamoreaux and his Los Alamos group~\cite{lamo} developed a solid
state technique that can improve the sensitivity to the electron
EDM by $10^4$ to reach $10^{-31}~e$~cm. By operating at a lower
temperature it is feasible to eventually reach a sensitivity of
$10^{-35}~e$~cm, an improvement of eight orders of magnitude over
the present sensitivity. The time scale for these is uncertain, as
it is tied to funding prospects. Semertzidis and his Brookhaven
group~\cite{seme} plan to trap muons in storage rings and increase
the sensitivity of their EDM measurement by five orders of
magnitude. A new measurement has been presented by the Sussex
group~\cite{hinds}.
A number of other experiments aim for an improvement in
sensitivity by one or two orders of magnitude, and involve nuclear
EDMs.

Improving the EDM measurements by two or more orders of magnitude
will test Split Supersymmetry as well as traditional
supersymmetric unified theories with low-energy supersymmetry. In
the latter, even if all the soft supersymmetry breaking terms are
real and consequently do not give rise to CP-violation, the normal
Kobayashi-Maskawa phase can be transmitted from the quark to the
lepton sector because of quark-lepton unification. In addition,
the large flavor breaking, mostly due to the large top mass,
results in non-universal slepton and squark mass
matrices~\cite{nonun}. As a result, these theories predict
electron and quark EDMs near the present limits. Improvement of
the sensitivity by two or more orders of magnitude will be a
significant test of these theories.

In Split Supersymmetry, CP violation in SM-fermion processes
occurs only at two loops. However, since charginos and neutralinos
have tree-level couplings with Higgs and gauge bosons, we expect
one-loop CP violation in processes involving $H$, $W$, $Z$ and
$\gamma$. One such process is the CP-violating Higgs decay into
two photons, generated by the operator in \eq{opcaz}. The rate is
\beq \frac{\Gamma (H\to \gamma \gamma )_{{\rm CP}^-}} {\Gamma
(H\to \gamma \gamma )_{{\rm CP}^+}}=\left( \frac{{\rm Im}~I} {{\rm
Re}~I} \right)^2, \eeq \beq I=\frac{4}{3} F_{1/2}\left(
\frac{4m_t^2}{m_H^2}\right) +F_{1}\left(
\frac{4M_W^2}{m_H^2}\right) + \frac{\sqrt{2}}{g}\sum_{i=1}^2\left(
\gtild^*U_{i2}V_{i1}+ \gtilu^*U_{i1}V_{i2}\right)
\frac{M_W}{m_{\chi^+_i}} F_{1/2}\left(
\frac{4m_{\chi^+_i}^2}{m_H^2}\right) , \label{cphis} \eeq \bea
F_{1/2}(x)&=&-2x\left[ 1+(1-x)\left( {\rm arcsin}
\frac{1}{\sqrt{x}}\right)^2
\right] , \\
F_{1}(x)&=&2+3x\left[ 1+(2-x)\left( {\rm arcsin}
\frac{1}{\sqrt{x}}\right)^2 \right] . \eea

\begin{figure}
\begin{center}
\epsfig{file=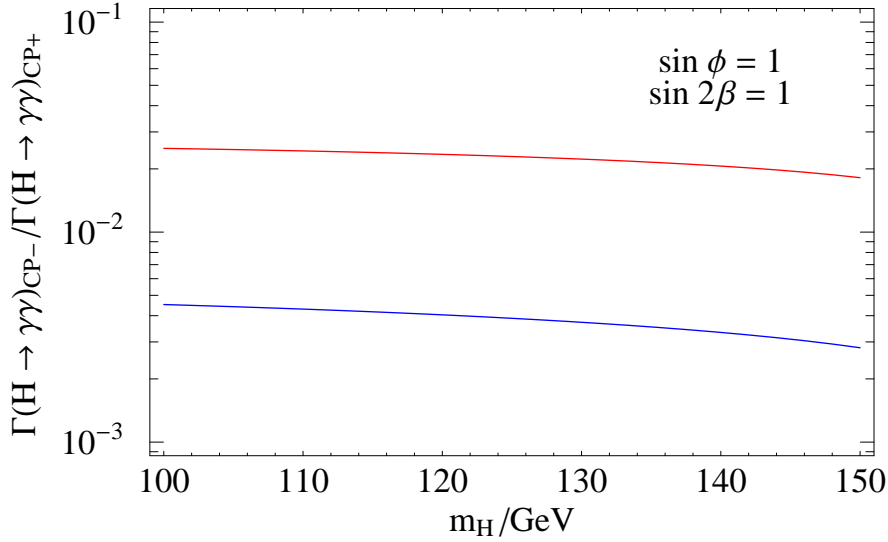,width=0.7\textwidth}
\end{center}
\caption{$\Gamma (H\to \gamma \gamma )_{{\rm CP}^-} /\Gamma (H\to
\gamma \gamma )_{{\rm CP}^+}$ as a function of the Higgs mass. We
take $\sin\Phi =1$ with $\Phi ={\rm arg}(\gtilu^*\gtild^* M_2\mu)$
and $\sin 2\beta =1$ using \eq{grel} at the chargino mass scale.
The top line corresponds to $m_{\chi^+_1}=100\gev$ and
$m_{\chi^+_2}=200\gev$, while the bottom line corresponds to
$m_{\chi^+_1}=150\gev$ and $m_{\chi^+_2}=300\gev$.} \label{figh}
\end{figure}

Using the same procedure we followed for the EDM, we can simplify
\eq{cphis} in the limit of heavy, but non-degenerate, charginos.  For
$m_H<2M_W$, we obtain 
\bea {\rm Im} I &=& \frac{4M_W^2 \gtilu \gtild
  M_2\mu \sin \Phi}{g^2(M_2^2- \mu^2)}\left[ M_2^{-2} F_{1/2}\left(
    \frac{4M_2^2}{m_H^2}\right) -\mu^{-2} F_{1/2}\left(
    \frac{4\mu^2}{m_H^2}\right) \right]
+{\cal O}\left( \frac{M_W^2}{M_2\mu}\right) \nonumber \\
&=& \frac{16M_W^2 \gtilu \gtild \sin \Phi}{3g^2M_2 \mu}+ {\cal
  O}\left( \frac{M_W^2}{M_2\mu}, \frac{m_H^2}{M_2\mu} \right) .
\label{struz} 
\eea 
In \fig{figh} we show the ratio $\Gamma (H\to \gamma \gamma )_{{\rm
    CP}^-} /\Gamma (H\to \gamma \gamma )_{{\rm CP}^+}$, which takes
interesting values for light charginos. The scaling with chargino
masses of the result shown in \fig{figh} can be easily read from
\eq{struz}.

Measurements of the polarization of the photons from Higgs decay
is experimentally challenging. However, the CP quantum number of
the Higgs boson can be measured at a $\gamma \gamma$
collider~\cite{gamcol}. The Higgs production amplitude in the
photon-fusion CP-even channel is proportional ${\vec{\epsilon}}_1
\cdot {\vec{\epsilon}}_2$, where ${\vec{\epsilon}}_{1,2}$ are the
polarization vectors of the two photon beams. On the other hand,
the amplitude in the CP-odd channel is proportional to
${\vec{\epsilon}}_1 \times {\vec{\epsilon}}_2 \cdot
{\vec{k}}_\gamma$, where ${\vec{k}}_\gamma$ is the photon
momentum. By varying the linear polarization of the colliding
photon beams, one can test the existence of the CP-odd
Higgs-photon couplings induced by the chargino loop. In Split
Supersymmetry the prospects are particularly favorable, since
phases of order one are presently allowed. Also notice that in
Split Supersymmetry we have a one-to-one relation between the
prediction of the EDM and of the CP-violating Higgs decay.

If charginos and neutralinos are produced at colliders,
experiments can directly test the presence of CP-violating phases.
One effect of these phases is to modify the relations between
model parameters and observables, even for CP-conserving
processes. Comparison between masses, cross sections and
distributions can single out the need to introduce phases for the
underlying parameters. However, it will be crucial to test the
phases in CP-violating processes. At a linear collider, one can
study asymmetries in the CP-odd spin-momentum product
${\vec{s}}_{\chi_i}\cdot {\vec{k}}_{e^-} \times
{\vec{k}}_{\chi_i}$ in $e^+e^-\to \chi_i \chi_j$ (for a recent
study see \cite{cpcol} and references therein). In Split
Supersymmetry, we can take advantage of the order-unity phases
allowed by EDM, but we lack the tree-level interference with
amplitudes induced by slepton exchange.

\section{Proton Decay}
\label{prosec}

In this section we discuss proton decay in Split Supersymmetry
with SU(5) unification.  The contribution from the model-dependent
dimension-5 operator is suppressed by four powers of the
supersymmetry-breaking scale $\tilde m$ and is subdominant for
$\tilde m \gsim 100\TeV$, if (as is commonly assumed) the
amplitude is suppressed by two light quark/lepton masses. In Split
Supersymmetry  the dimension-5 $p$-decay operator does not need to
be suppressed by light fermion masses, since the heavy  squarks can
provide an adequate suppression. This can allow for new theoretical
possibilities. For example, Yukawa couplings (flavor) can
originate in the right-handed fermion sector and there will be no
mass  suppressions associated with the dimension-5 operator
$qqq\ell$, constructed out of only left-handed fermions.  In such
theories the dimension-5  $p$-decay would depend on the masses of
left-handed squarks and gaugino/higgsinos, and would be highly
model-dependent. We now focus on the dimension-6 operator
contribution from heavy gauge-boson exchange, which is model
independent.

The dominant decay mode is $p\rightarrow e^+\pi^0$ and the decay
width is~\cite{Langacker:1980js}
\begin{equation}
   \label{eq:width}
\tau_p^{-1}\equiv \Gamma(p\rightarrow e^+\pi^0) = \frac{\pi\,
m_p}{4f^2_\pi}
   \alpha_N^2(1+D+F)^2\left(\frac{\alpha_{\rm{GUT}}}{M^2_V}\right)^2
   \left[ A^2_R +A^2_L\left( 1+|V_{ud}|^2\right)^2\right] \,,
\end{equation}
where $M_V$ is the GUT gauge-boson mass and $\alpha_{\rm{GUT}}$ is
the unification coupling. Also, $f_\pi=131\MeV$ is the pion decay
constant, $(1+D+F)=2.25$ is the chiral Lagrangian factor and we
use the lattice result $\alpha_N = 0.015\GeV^3$ for the hadronic
matrix element parameter $\alpha_N$~\cite{Aoki:1999tw}. Note that
the lattice determination is 3--5 times larger than the smallest
estimate among QCD model calculations. The factors $A_L$, $A_R$
account for the leading-log renormalization of the operators
${\cal O}_L=(u_c^\dagger \sigma_\mu q_1)(e_c^\dagger \sigma^\mu
q_1)$, ${\cal O}_R=(u_c^\dagger \sigma_\mu q_1)(d_c^\dagger
\sigma^\mu \ell_1)$ (in Weyl notation). They can be decomposed
into four parts,
\begin{equation}
   \label{eq:A}
   A_{L,R} = A_{\rm{QCD}}\, A^{L,R}_{\rm SM}\, A^{L,R}_{\rm
   SSSM}\, A^{L,R}_{\rm MSSM} \,.
\end{equation}
The factor $A_{\rm QCD}$ accounts for the QCD
renormalization-group evolution below $M_Z$. Since the matrix
elements are evaluated at the scale $Q=2.3\GeV$, we have
\begin{equation}
   \label{eq:AQCD}
   A_{\rm QCD} = \left[ \frac{\alpha_3(m_b)}{\alpha_3(M_Z)}
   \right]^{\frac{6}{23}} \left[ \frac{\alpha_3(Q)}{\alpha_3(m_b)}
   \right]^{\frac{6}{25}} \simeq 1.25 \,.
\end{equation}
The SM running up to the $R$-breaking scale (gaugino and higgsino
masses) and the subsequent running up to the supersymmetry
breaking scale $\mtil$ give~\cite{Buras:1977yy} \bea
   A^L_{\rm SM, SSSM} &=& \left[ \frac{\alpha_3(\tilde
       m)}{\alpha_3(M_Z)} \right]^{\frac{2}{b_3}} \left[
     \frac{\alpha_2(\tilde m)}{\alpha_2(M_Z)} \right]^{\frac{9}{4b_2}}
   \left[ \frac{\alpha_1(\tilde m)}{\alpha_1(M_Z)}
   \right]^{\frac{23}{20b_1}} \\
   A^R_{\rm SM, SSSM} &=& \left[ \frac{\alpha_3(\tilde
       m)}{\alpha_3(M_Z)} \right]^{\frac{2}{b_3}} \left[
     \frac{\alpha_2(\tilde m)}{\alpha_2(M_Z)} \right]^{\frac{9}{4b_2}}
   \left[ \frac{\alpha_1(\tilde m)}{\alpha_1(M_Z)}
   \right]^{\frac{11}{20b_1}} \,.
\eea Here $b_i$ are the $\beta$-function coefficients $b_i=-8\pi^2
dg_i^{-}2/d\ln \mu$ and
$(b_1,b_2,b_3)_{\rm SM} = (41/10,-19/6,-7)$ and
$(b_1,b_2,b_3)_{\rm SSSM} = (9/2,-7/6,-5)$.  Above the
supersymmetry breaking scale $\tilde m$, the 4-fermion operators
mix with their supersymmetric analogs involving scalars,
giving~\cite{Ibanez:1984ni} (the U(1) coefficients have been
calculated in ref.~\cite{Munoz:1986kq}) \bea
   A^L_{\rm MSSM} & =& \left[ \frac{\alpha_3(M_{\rm
         GUT})}{\alpha_3(\tilde m)} \right]^{\frac{4}{3b_3}} \left[
     \frac{\alpha_2(M_{\rm GUT})}{\alpha_2(\tilde m)}
   \right]^{\frac{3}{2b_3}} \left[ \frac{\alpha_1(M_{\rm
         GUT})}{\alpha_1(\tilde m)}
   \right]^{\frac{23}{30b_1}} \\
   A^R_{\rm MSSM} & = &\left[ \frac{\alpha_3(M_{\rm
         GUT})}{\alpha_3(\tilde m)} \right]^{\frac{4}{3b_3}} \left[
     \frac{\alpha_2(M_{\rm GUT})}{\alpha_2(\tilde m)}
   \right]^{\frac{3}{2b_3}} \left[ \frac{\alpha_1(M_{\rm
         GUT})}{\alpha_1(\tilde m)} \right]^{\frac{11}{30b_1}} \,,
\eea with $(b_1,b_2,b_3)_{\rm MSSM} = (33/5,1,-3)$. The
expressions above can be further factorized in order to take into
account the thresholds at the EW, $R$, supersymmetry and SU(5) breaking
scales. The appropriate $\beta$-function coefficients can be found
in Table~\ref{tab:beta}.

\begin{figure}
\begin{center}
\epsfig{file=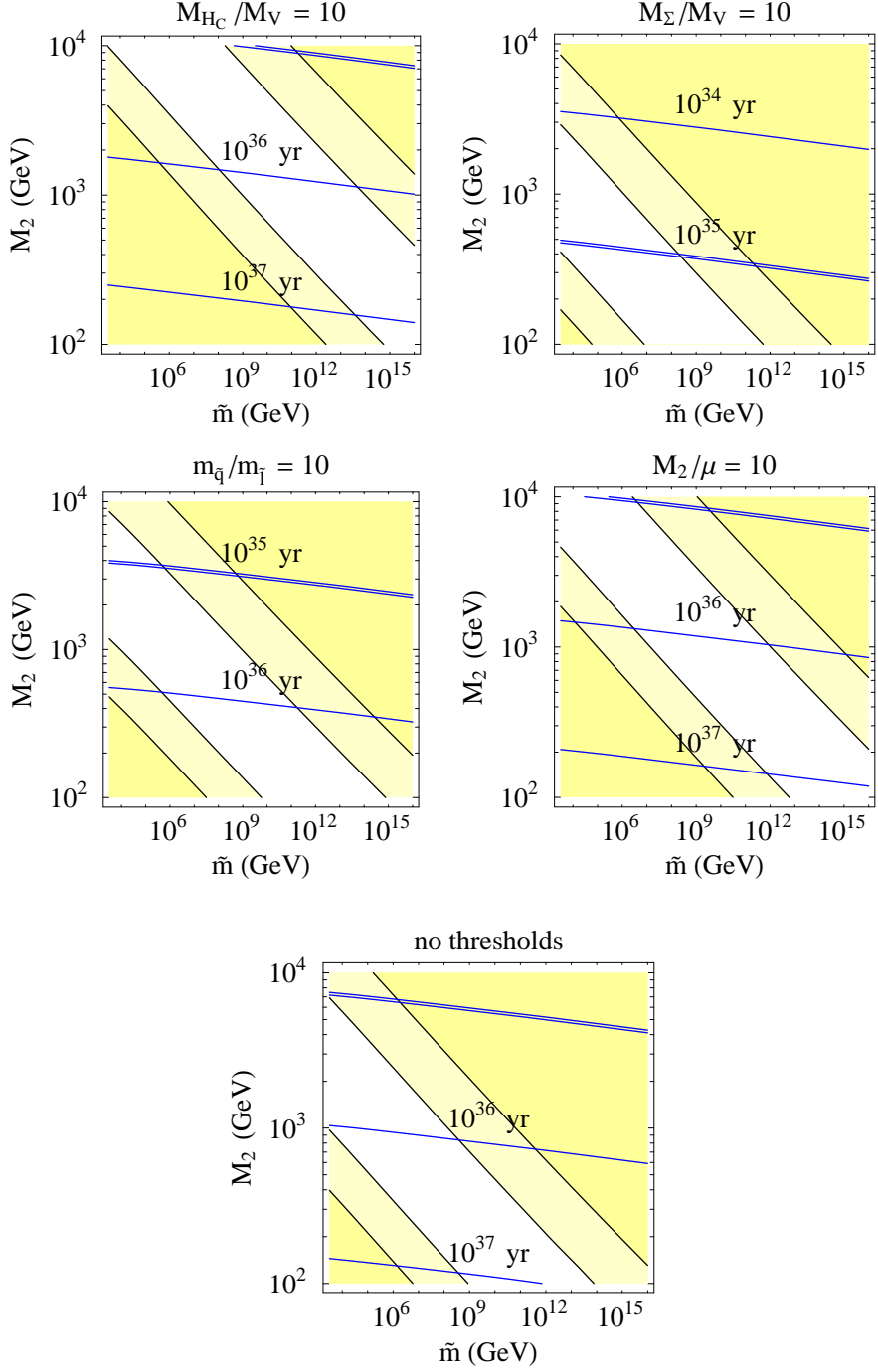,width=0.7\textwidth}
\end{center}
\caption{Contours of the proton-decay lifetime, with the thicker
line corresponding to $\tau_p = 10^{35}\,{\rm yr}$. The diagonal
bands represent the $1\sigma$ and $2\sigma$
   constraints corresponding to $\alpha_3(M_Z) = 0.119\pm0.003$.
The 4 different thresholds that appear in
   eqs.~(\ref{eq:shifts1})--(\ref{eq:shifts2}) are included as
indicated.} \label{fig:pdecay}
\end{figure}

The proton decay rate crucially depends on the GUT gauge-boson mass
$M_V$ and on the unified gauge coupling $\alpha_{\rm GUT}$ at that
scale. The two-loop determination of those quantities can be found in
ref.~\cite{noi}. The corresponding prediction for the proton lifetime
is shown in fig.~\ref{fig:pdecay}, lowest panel, as a contour plot in
the $\tilde m$-$M_2$ plane. We have used $\mu = M_2$ and we have
obtained $M_3$ from $M_2(M_{\rm GUT}) = M_3(M_{\rm GUT})$.  The
$1\sigma$ and $2\sigma$ constraints from $\alpha_3(M_Z) =
0.119\pm0.003$ are also shown. Possible GUT thresholds have been
neglected.

The prediction in fig.~\ref{fig:pdecay} can be compared with the
present 90\% CL limit from SK, $\tau(p\rightarrow e^+\pi^0) >
5.0\times 10^{33}\,{\rm yr}$~\cite{Suzuki:2001rb}, with the
expected sensitivity after 10 SK live-years, $\tau(p\rightarrow
e^+\pi^0) \gsim 10^{34}\,{\rm yr}$, and with the expected
sensitivity at the next generation of water Cherenkov
experiments~\cite{Itow:2001ee,Jung:1999jq,Diwan:2003uw}. For
example, a 650 kton UNO-like experiment could reach a sensitivity
of $1.6\times 10^{35}\,{\rm yr}$ in 10 years of
running~\cite{Rubbia:2004yq}.

The dependence of the $p$-decay rate on $\tilde m$ is quite mild.
Actually, at one loop, $M_{\rm GUT}$ depends on $\tilde m$ only
through the heavy Higgs contribution to the $\beta$-functions.
This has the effect of decreasing $M_{\rm GUT}$~\cite{noi} and
therefore of enhancing the $p$-decay rate.  However, the effect of
$M_{\rm GUT}$ is partially compensated by the smaller values of
$\alpha_{\rm GUT}$ found in Split Supersymmetry~\cite{noi}.
Notice that additional matter in complete SU(5) representations at
intermediate scales has no effect on $M_{\rm GUT}$, but increases
$\alpha_{\rm GUT}$. Therefore, a shorter proton lifetime will be
found, for instance, in models with intermediate messenger
particles, as in gauge mediation or in models with direct
mediation, as discussed in sect.~\ref{direct}. We will comment on
this possibility at the end of this section.

The proton lifetime becomes shorter as the value of $M_2$
increases. Therefore proton-decay searches can be viewed as
complementary to direct searches at accelerators, although they
have a rather modest reach in parameter space. 
% For illustrative
% purposes, in fig.~\ref{fig:pdecay} we are actually showing values
% of $M_2$ beyond what is allowed by the dark-matter constraint.
This figure shows that a correct value of $\alpha_3$ implies an
upper limit on $M_2$ and a lower limit on $\tau_p$.  As a
consequence, larger values of $M_2$ and shorter proton lifetimes
are allowed for lower values of $\tilde m$.

We now investigate the effect of three types of thresholds
corrections. \\[1mm]
$\bullet$ $R$-breaking scale thresholds due to $\mu\neq M_2$. \\[1mm]
$\bullet$ supersymmetry-breaking scale thresholds associated to a
non-degeneracy among the sfermion masses or between the sfermion
masses and the heavy Higgs mass $m_H$ (the latter will turn out
not to play a role at 1 loop).  For simplicity, we will assume
that all squarks and all sleptons are separately degenerate with
common masses $\msq$,
$\msl$ respectively. \\[1mm]
$\bullet$ SU(5)-breaking scale thresholds. We will consider the
effect of the heavy SU(5) vector superfields with mass $M_V$, of
the color triplet partners of the Higgs doublets with mass $\MHc$,
and of an adjoint SU(5)-breaking chiral superfield with mass
$\MAd$.

The one-loop thresholds can be incorporated in the 2-loop
evolution of the gauge couplings by using the following matching
conditions at the scales $M_V$, $m_H$ and $M_2$: \bea
   \frac{1}{\alpha_i(M_V-)} &=& \frac{1}{\alpha_i(M_V+)}
   +\frac{b^{H_C}_i}{2\pi} \ln\frac{M_V}{\MHc} +
   \frac{b^{\Sigma}_i}{2\pi} \ln\frac{M_V}{\MAd} \nonumber
\\
   \frac{1}{\alpha_i(m_H-)} &=& \frac{1}{\alpha_i(m_H+)}
   +\frac{b^{\tilde q}_i}{2\pi} \ln\frac{\msl}{\msq} +
   \frac{b^{\tilde f}_i}{2\pi} \ln\frac{m_H}{\msl}
\label{eq:thresholds}
\\
   \frac{1}{\alpha_i(M_2-)} &=& \frac{1}{\alpha_i(M_2+)}
   +\frac{b^{\tilde H}_i}{2\pi} \ln\frac{M_2}{\mu} \,. \nonumber
\eea Here, $b^x_i$ represents the contribution of the particle $x$
to the $\beta$-function coefficient of the gauge coupling $g_i$.
The numerical values are listed in Table~\ref{tab:beta}.  The
coefficients $b_{\tilde f} = (2,2,2)$ are associated to the whole
set of sfermions, in a full SU(5) representation. Therefore, the
$\ln(m_H/\msl)$ term does not enter the determination of $M_V$ and
$\alpha_3$ at one loop.

\begin{table}
\centering
\begin{tabular}{|c||c|c|c||c|c|c||c|c|c|}
\hline $x$ & $\tilde H$ & $\tilde W$ & $\tilde G$ & $H$ & $\tilde
q$ &
$\tilde l$ & $V$ & $H_C$ & $\Sigma$ \\
\hline
$b^x_1$ & 2/5 & 0 & 0 & 1/10 & 11/10 & 9/10 & -10 & 2/5 & 0 \\
\hline
$b^x_2$ & 2/3 & 4/3 & 0 & 1/6 & 3/2 & 1/2 & -6 & 0 & 2 \\
\hline
$b^x_3$ & 0 & 0 & 2 & 0 & 2 & 0 & -4 & 1 & 3 \\
\hline
\end{tabular}
\caption{Contributions to the $\beta$-function coefficients from
particle $x$. The $\beta$-function coefficients for a theory with
the particle content of SM $+\{x\}$ (where $\{x\}$ describes the
set of new particles) are obtained by the relation $b_i=b_i^{\rm
SM} +\sum_x b_i^x$, where $b_i^{\rm SM}=(41/10,-19/6,-7)$ and
$b_i^x$ are given in the table.} \label{tab:beta}
\end{table}

An analytic understanding of the threshold effects on the
determination of $M_{\rm GUT}$ and $\alpha_3$ can be easily gained
in the one loop approximation,
%% \label{eq:shifts}
%% \bea
%%   \ln\frac{M_V}{(M_V)_0}& =&
%%   -\frac{1}{14} \ln\frac{M_V}{\MHc}
%%   +\frac{5}{14} \ln\frac{M_V}{\MAd}
%%   +\frac{1}{14} \ln\frac{\msl}{\msq}
%%   +\frac{1}{21} \ln\frac{M_2}{\mu}
%% \label{eq:shifts1}
%%  \\
%%   \frac{1}{\alpha_3}& =& \left( \frac{1}{\alpha_3} \right)_0
%%   +\frac{1}{2\pi} \left(
%%     \frac{9}{7} \ln\frac{M_V}{\MHc}
%%     -\frac{3}{7} \ln\frac{M_V}{\MAd}
%%     +\frac{3}{14} \ln\frac{\msl}{\msq}
%%     -\frac{6}{7} \ln\frac{M_2}{\mu} \right) \,,
%% \label{eq:shifts2}
%% \\
%% \frac{1}{\alpha_{\rm GUT}} &=&
%% \left( \frac{1}{\alpha_{\rm GUT}} \right)_0 -\frac{1}{2\pi}
%% \left(
%% -\frac{1}{14} \log\frac{M_V}{\MHc}
%% +\frac{33}{14} \log\frac{M_V}{\MAd}
%% +\frac{11}{7} \log\frac{\msl}{\msq}
%% +\frac{5}{7} \log\frac{M_2}{\mu}
%% +2 \log\frac{m_H}{\msl}
%% \right)
%% \label{eq:shifts3}
%% \eea
\label{eq:shifts} \bea
   \delta \ln M_V & =&
   -\frac{1}{14} \ln\frac{M_V}{\MHc}
   +\frac{5}{14} \ln\frac{M_V}{\MAd}
   +\frac{1}{14} \ln\frac{\msl}{\msq}
   +\frac{1}{21} \ln\frac{M_2}{\mu}
\label{eq:shifts1} \nonumber
  \\
   \delta\alpha_3^{-1}& =&
   \frac{1}{2\pi} \left(
     \frac{9}{7} \ln\frac{M_V}{\MHc}
     -\frac{3}{7} \ln\frac{M_V}{\MAd}
     +\frac{3}{14} \ln\frac{\msl}{\msq}
     -\frac{6}{7} \ln\frac{M_2}{\mu} \right) \,,
\label{eq:shifts2}
\\
\delta\alpha_{\rm GUT}^{-1} &=& -\frac{1}{2\pi} \left(
-\frac{1}{14} \log\frac{M_V}{\MHc} +\frac{33}{14}
\log\frac{M_V}{\MAd} +\frac{11}{7} \log\frac{\msl}{\msq}
+\frac{5}{7} \log\frac{M_2}{\mu} +2 \log\frac{m_H}{\msl} \right)
\label{eq:shifts3} \nonumber \eea where $\delta x$ denotes the
contribution to $x$ due to the thresholds. A positive contribution
$\delta \alpha_3^{-1}$ lowers the $\alpha_3$ bands in
fig.~\ref{fig:pdecay}, while a positive contribution $\delta \ln
M_V$ rises the lifetime contours. This result is evident
in the figures, which however contains the full 2-loop effects.

From the first panel of fig.~\ref{fig:pdecay} we observe that the
Higgs triplet has a significant effect on $\alpha_3$, and a limited
effect on $\tau_p$. For heavy $H_C$, it is possible to obtain
$\tau_p<10^{35}$~yr, but only at the price of very heavy gauginos.
The second panel of fig.~\ref{fig:pdecay} shows that $\Sigma$ has a
considerable effect on $\tau_p$, but not on $\alpha_3$. For a heavy
$\Sigma$, proton decay is well within the reach of a megaton water
Cherenkov experiment, even for gaugino masses accessible to the LHC
and CLIC. The third panel shows that having squarks heavier than
sleptons improves the chances for proton-decay searches. We should
remark however that squarks much heavier than sleptons is not
consistent with the simplest unification relations. The fourth panel
illustrates how a threshold with higgsinos lighter than gauginos
modifies the $\alpha_3$ prediction, allowing for larger proton-decay
rates. The latter plot illustrates a particularly interesting and
fairly generic case where the $p$-decay rate may be observable and
consistent with both unification and DM abundance, for a broad range
of scalar masses. It corresponds to $M_2 \sim 10\,{\rm TeV}$ and $\mu
\sim 1 \,{\rm TeV}$. This is the situation we have if $\mu$ starts out
much smaller than the gaugino masses at the UV scale. Then, the
physical gaugino masses will be roughly an order of magnitude heavier
than the low-energy value of $\mu$ \cite{noi}, leading to a
mostly-higgsino LSP -- which must then weigh about a TeV to be the DM,
leading to the parameter choice of the 
fourth plot of fig.~\ref{fig:pdecay}. This suggests
that potentially observable $p$-decay will be a fairly common
characteristic of theories which start out with a small $\mu$ in the
UV. On the other hand, it will be hard to see such a $\sim$TeV-mass
Higgsino at the LHC.

Finally, we consider the effect of additional matter in complete
vectorial SU(5) multiplets. Such extra multiplets arise in simple
implementations of direct mediation, see sect.~\ref{direct}. The
leading effect, which we calculate, is due to the increase in
$\alpha_{\rm GUT}$. Note that the extra multiplets also affect the
operator renormalization coefficients $A_{L,R}$ at one loop and
the determination of $M_{\rm GUT}$ at two loops. The effect on
$\alpha_{\rm GUT}$ is given by
\begin{equation}
   \label{eq:extra}
   \frac{1}{\alpha_{\rm GUT}} = \left(\frac{1}{\alpha_{\rm GUT}}
   \right)_0 -\frac{N}{2\pi} \log\frac{M_{\rm GUT}}{\tilde m} \,,
\end{equation}
where the index ``0'' refers to the determination of $\alpha_{\rm
  GUT}$ in the absence of extra matter, which corresponds to the predictions
shown in Fig.~\ref{fig:pdecay}, and $N$ is the Dynkin index of the
gauge representation of the extra matter ($N = 1$ for a single $5+\bar
5$, $N = 3$ for a single $10+\bar {10}$). In this simple approximation,
the perturbativity limit on $N$ is about 15 for $\tilde m =
10^{11}\GeV$ and scales with $1/\log(M_{\rm GUT}/\tilde m)$.  The
limit on $N$ from proton decay is approximately the same.  If the
Dynkin index $N$ is sizeable but not fine-tuned to be close to the
perturbativity limit, we can expect an increase of $\alpha_{\rm GUT}$
by a factor ${\cal O}(1)$, which leads to an increase of the proton
decay rate by a factor ${\cal O}(1)^2$.

In conclusion, some versions of Split Supersymmetry, such as those with
higgsino LSP, naturally allow for the possibility of potentially
observable proton lifetime, consistent with unification and DM.

%In conclusion, from the point of view of proton-decay searches, Split
%Supersymmetry does not significantly change the ordinary
%supersymmetric prediction from dimension-6 operators~\cite{protd},
%unless new particles in complete GUT multiplets are added at
%intermediate scales. However, since we have abandoned the naturalness
%criterion, we are allowed to consider heavier gauginos and higgsinos.
%This has the effect to enhance the proton-decay rate, consistently
%with gauge-coupling unification.

\section{Conclusions}
\label{concsec}

In this paper we have investigated theoretical, cosmological and
phenomenological aspects of Split Supersymmetry.

We have shown that the mass spectrum of Split Supersymmetry, which
was originally motivated by low-energy considerations (dark matter
and prediction of $\alpha_s$), can actually have a natural
justification in the high-energy theory. An approximate
$R$-symmetry, which forbids gaugino and higgsino mass terms, can
explain the hierarchies of the mass spectrum of Split
Supersymmetry. Nevertheless, Split Supersymmetry can emerge from
the high-energy theory not necessarily as a consequence of an
imposed $R$-symmetry, but simply as a result of the pattern of
supersymmetry breaking. Using the property that $R$-symmetric soft
terms correspond to dimension-2 operators and $R$-breaking soft
terms to dimension-3 operators, we have shown that supersymmetry
$D$-breaking leads to Split Supersymmetry, while $F$-breaking
leads to the usual mass spectrum with no hierarchies among
sparticles. In the case of $D$-breaking, the underlying
$R$-symmetry protecting higgsino and gaugino masses emerges as an
accidental symmetry, much alike the approximate lepton-number
conservation that protects neutrino masses in the Standard Model.

We have also studied the general structure of Split Supersymmetry
as we vary the mechanism of supersymmetry breaking or, ultimately,
the gravitino mass $m_{3/2}$. Since the squark and slepton mass
scale $\mtil$ is an arbitrary parameter, no longer tied to the
weak scale, the theory offers novel scenarios with different
theoretical, cosmological and phenomenological implications. In
general, the different features of Split Supersymmetry are best
parametrized by the values of  $m_{3/2}$ and $\mtil$.

Whenever $m_{3/2}$ is much larger than the weak scale (which, in
particular, is the case for gravity mediation with $\mtil
\gg$~TeV), then anomaly-mediated contribution can upset the
hierarchy between supersymmetric scalars and fermions. This issue
has been addressed in sect.~\ref{anomed}. We have found that in
theories where supersymmetry breaking is intimately tied to gravity, such
that supersymmetry is restored in the $\mpl \to \infty$ limit, the 
anomaly-mediated contributions are suppressed and there is no obstacle to
considering $m_{3/2}$ far above the TeV scale.

We have also discussed an interesting option, available in Split Supersymmetry
but not in low-energy supersymmetry, to break supersymmetry and directly communicate
it to squarks and sleptons at tree-level, via renormalizable couplings.
These theories naturally generate a large hierarchy between
the scalars and the gauginos/Higgsinos, and also permit an
interesting spectrum with a gravitino LSP. Such models inevitably
require extra charged matter at intermediate scales, raising the
value of the unified gauge coupling at the GUT scale, and possible
enhancing the proton decay rate from dimension 6 operators to
observable levels.

In Split Supersymmetry, the gravitino mass can vary in a much
larger range than in ordinary low-energy supersymmetry and this
has new cosmological implications. In sect.~\ref{grasec} we have
revisited gravitino production in the early universe. The decay of
thermalized heavy particles with mass $\mtil$ plays a new and
important role, creating a gravitino abundance which is
independent of the reheat temperature $T_R$, and which depends
only on physical particle masses.

Heavy gravitinos eventually decay into supersymmetric particles. A
peculiarity of gravitinos with masses much larger than the weak
scale (and thus a peculiarity of Split Supersymmetry) is that the
decay occurs well before the beginning of nucleosynthesis, and
therefore the gravitino abundance before decay is not tightly
constrained. If the decay happens after neutralinos have decoupled
from the thermal bath, then a non-thermal population of
neutralinos is generated. We have studied the range of parameters
for which this non-thermal $\chi$ component can account for the
observed dark-matter.

The existence of a non-thermal neutralino population has important
phenomenological consequences. The dark-matter argument is a
critical ingredient of Split Supersymmetry, providing a tight
constraint on the higgsino and gaugino parameters. The non-thermal
$\chi$ density gives us more freedom, allowing us to consider
neutralino parameters that are normally excluded, because they
lead to an insufficient thermal density. In spite of this freedom,
the upper bound on $m_\chi$, derived from thermal dark-matter
considerations, is not evaded, but actually reinforced. This is
essential, as the bound on $m_\chi$ provides the necessary link
between gaugino-higgsino masses and the weak scale.

If the scale of supersymmetry mediation is much smaller than
$\mpl$, the gravitino can be light, and it can even become the
LSP. In this case, the neutralino can no longer form the dark
matter, and the connection between its mass and the weak scale
disappears (although an upper bound on its mass still exists,
under the condition that gravitinos constitute the dark matter). A
gravitino LSP is a viable dark-matter candidate. The collider
signals of this case are also quite distinctive, whenever the
decays into gravitinos occur inside the detector. A peculiar
signature of Split Supersymmetry with light gravitinos is the
gluino decay ${\tilde g}\to {\tilde G} g$. This decay can dominate
over the ordinary 3-body process into neutralinos and charginos,
when the scale of supersymmetry-breaking mediation is close to
$\mtil$. This is the case of direct mediation.

Next, we have addressed some phenomenological issues. Split
Supersymmetry solves the flavour and CP problem, by considering
very heavy squarks and sleptons. However, the effective theory
below the scale $\mtil$ contains a CP violating phase. This leads
to predictions for electron and neutron EDMs, which are just
beyond the present experimental limits, for a phase of order unity
and for weak-scale chargino masses. The difference with respect to
ordinary low-energy supersymmetry, where phases have to be smaller
than about $10^{-2}$ to be consistent with present limits, is that
EDM are only generated at two loops in Split Supersymmetry.
Because of the intense experimental activity aiming at improving
the EDM measurements, our prediction offers the exciting
possibility that Split Supersymmetry could start revealing itself
even before the LHC is operational. Other tests of CP violation at
colliders are also possible, as discussed in sect.~\ref{edmsec}.

Finally, we have investigated how Split Supersymmetry affects
proton decay. Dimension-5 operators are naturally suppressed,
because of the heavy squarks. Since Split Supersymmetry has the
effect of reducing the value of the unification mass, proton-decay
dimension-6 operators can be enhanced. However, in the minimal
version of the model, the unification coupling constant increases,
nearly compensating the effect of a lower $M_{GUT}$. Low-energy
thresholds can enhance the proton-decay rate, and this is
especially true if we consider very heavy higgsinos and gauginos.
In this sense, searches for proton decay can be complementary to
collider searches, since they are most sensitive to a region of
sparticle masses hard to test at colliders.

In conclusion, we have shown that Split Supersymmetry can
naturally emerge from a fundamental high-energy theory. It has
specific implications for the origin of the dark matter and it
makes the exciting prediction that EDMs are naturally within the
reach of ongoing experiments.

\bigskip

We acknowledge communications and discussions with D.~DeMille,
H.~Haber, L.~Iba\~nez, S.~Lamoreaux, M.~Luty,
R.~Rattazzi, A.~Ritz, Y.~Semertzidis and
E.~Witten.

%%%%%%%%%%%%%%%%%%%%%%%%%%%%%%%%%%%%%%%%%%%%%%%%%%%%%%%%%%%%%%%%%%%%%%%%

\end{document}